\documentclass[12pt]{article} 

\usepackage{amssymb}
\usepackage{ulem, makecell, multirow}
\usepackage{booktabs}
\usepackage{graphicx, subcaption}
\usepackage{xr}
\usepackage{color}
\newcommand{\real}[1]{{\mathbb R}^{#1}}
\newcommand{\U}{{\mathbf U}}

\newcommand{\W}{{\mathbf W}}

\newcommand{\A}{{\mathbf A}}

\newcommand{\B}{{\mathbf B}}
\newcommand{\Q}{{\mathbf Q}}
 \newcommand{\Yhat}{\widehat{\mathbf Y}}

\usepackage{float}
\usepackage{enumerate}

 \usepackage[semicolon]{natbib}
\usepackage{amsmath}
\usepackage{amsfonts}
\usepackage{amssymb}
\usepackage{latexsym}
\usepackage{verbatim}
\usepackage{times}
\usepackage{color}
\usepackage{graphicx}
\usepackage{url}
\usepackage{multirow}
\newcommand{\zerobf}{0}
\newcommand{\onebf}{1}

\newcommand{\boxedeqn}[1]{%
  \[\fbox{%
      \addtolength{\linewidth}{-2\fboxsep}%
      \addtolength{\linewidth}{-2\fboxrule}%
      \begin{minipage}{\linewidth}%
      \begin{equatxion}#1\end{equation}%
      \end{minipage}%
    }\]%
}

\usepackage[left]{lineno}
\usepackage{setspace}
\newtheorem{prop}{Proposition}[section]

\newtheorem{example}{Example}

\numberwithin{equation}{section}
\numberwithin{figure}{section}

\usepackage[margin=1in]{geometry}

\newcommand{\indep}{\;\, \rule[0em]{.03em}{.67em} \hspace{-.25em}
\rule[0em]{.65em}{.03em} \hspace{-.25em}
\rule[0em]{.03em}{.67em}\;\,}

\newcommand{\spn}{\mathrm{span}}
\newcommand{\diag}{\mathrm{diag}}

\newcommand{\var}{\mathrm{var}}

\newcommand{\rank}{\mathrm{rank}}

\newcommand{\vecc}{\mathrm{vec}}

\newcommand{\I}{\mathbf I}
 
  \newcommand{\Ubf}{\mathbf U}

\newcommand{\R}{{\mathbf R}}
\newcommand{\avar}{\mathrm{avar}}

\newcommand{\espc}{{\mathcal E}}
\newcommand{\Ybar}{\bar{\Y}}
\newcommand{\betabfhat}{\widehat{\greekbold{\beta}}}
 \newcommand{\Cm}{\mathrm{cm}}
\newcommand{\Em}{\mathrm{em}}

\newcommand{\Um}{\mathrm{um}}
\newcommand{\Ecm}{\mathrm{ecm}}

\newcommand{\Sem}{\mathrm{sem}}
\newcommand{\Secm}{\mathrm{secm}}
\newcommand{\X}{{\mathbf X}}
\newcommand{\x}{{\mathbf x}}

\newcommand{\Y}{{\mathbf Y}}

\newcommand{\abf}{\mathbf a}

\newcommand{\bb}{{\mathbf b}}

 \newcommand{\Gammabfhats}{\widehat{\greekbold{\scriptstyle \Gamma}}}

 \newcommand{\1}1

\newcommand{\Xbar}{\bar{\X}}

\newcommand{\Z}{{\mathbf Z}}

\newcommand{\D}{\mathbf D}

\newcommand{\K}{\mathbf K}

\newcommand{\Pbf}{{\mathbf P}}

\newcommand{\T}{\mathbf T}

\newcommand{\V}{{\mathbf V}}

\newcommand{\bu}{{\mathbf u}}

\newcommand{\Lhat}{\hat{L}}

\newcommand{\M}{{\mathbf M}}

\newcommand{\Sbf}{{\mathbf S}}
\newcommand{\ba}{{\mathbf a}}

\newcommand{\cbf}{{\mathbf c}}

\newcommand{\C}{{\mathbf C}}

\newcommand{\e}{{\mathbf e}}

\newcommand{\G}{{\mathbf G}}
\newcommand{\Ghat}{\widehat{\G}}

\newcommand{\greekbold}[1]{\mbox{\boldmath $#1$}}
\newcommand{\phibf}{\greekbold{\phi}}
\newcommand{\alphabf}{\greekbold{\alpha}}
\newcommand{\alphabfhat}{\widehat{\alphabf}}

\newcommand{\etabf}{\greekbold{\eta}}

\newcommand{\betabf}{\greekbold{\beta}}

\newcommand{\betahat}{\widehat{\greekbold{\beta}}}

\newcommand{\Lambdabf}{\greekbold{\Lambda}}

\newcommand{\varepsilonbf}{\greekbold{\varepsilon}}

\newcommand{\Gammabf}{\greekbold{\Gamma}}

\newcommand{\Gammabfs}{{\greekbold{\scriptstyle \Gamma}}}
\newcommand{\Gammabfhat}{\widehat{\greekbold{\Gamma}}}

\newcommand{\Thetabf}{\greekbold{\Theta}}

\newcommand{\omegabf}{\greekbold{\omega}}
\newcommand{\Omegabf}{\greekbold{\Omega}}
\newcommand{\Omegabfhat}{\greekbold{\widehat{\Omega}}}
\newcommand{\Phibf}{\greekbold{\phi}}

\newcommand{\Sigmabf}{\greekbold{\Sigma}}
\newcommand{\Sigmabfhat}{\greekbold{\widehat{\Sigma}}}

\newcommand{\Sigmabfs}{{\greekbold{\scriptstyle \Sigma}}}

\newcommand{\spc}{{\mathcal S}}

\newcommand{\bspc}{{\mathcal B}}

 \newcommand{\aspc}{{\mathcal A}}
\newcommand{\uspc}{{\mathcal U}}
\newcommand{\gspc}{{\mathcal G}}

 \newcommand{\dspc}{{\mathcal D}}
\newcommand{\Phibfs}{\greekbold{\scriptstyle \Phi}}
\newcommand{\etabfhat}{\widehat{\etabf}}
 \newcommand{\phibfhat}{\widehat{\greekbold{\phi}}}

\newcommand{\bdiag}{\text{bdiag}}
\newcommand{\new}{\mathrm{new}}
\newcommand{\Obf}{\mathbf O}
\newcommand{\Lambdabfhat}{\widehat{\greekbold{\Lambda}}}

\newcommand{\Phibfhat}{\widehat{\Phibf}}

\newcommand{\colrank}{\text{colrank}}
\newcommand{\spant}{\text{span}}

\newcommand{\Thetabfhat}{\widehat{\greekbold{ \Theta}}}
\newcommand{\Thetabfhats}{\widehat{\greekbold{\scriptstyle \Theta}}}
\newcommand{\Phibfhats}{\widehat{\greekbold{\scriptstyle{\Phi}}}}

\newcommand{\Lambdabfs}{\greekbold{\scriptstyle \Lambda}}

\begin{document}

\title{Envelopes for multivariate linear regression with linearly constrained coefficients}

 \author{R. Dennis Cook,\thanks{{\small R. Dennis Cook is Professor, School of Statistics, University
    of Minnesota, Minneapolis, MN 55455 (E-mail: dennis@stat.umn.edu).}} 
    \; Liliana Forzani%
     \thanks{{\small Liliana Forzani is Professor, Facultad de Ingenier\'ia Qu\'imica, UNL. Researcher
      of CONICET, Santa Fe, Argentina (E-mail: liliana.forzani@gmail.com). }}
      \; and Lan Liu\thanks{{\small Lan Liu is Assistant Professor, School of Statistics, University
    of Minnesota, Minneapolis, MN 55455 (E-mail: liu1815@gmail.com).}} 
       }

\maketitle

\begin{abstract}
{A constrained multivariate linear model is a multivariate linear model with the columns of its coefficient matrix constrained to lie in a known subspace. This class of models includes those typically used to study growth curves and longitudinal data. Envelope methods have been proposed to improve estimation efficiency in the class of unconstrained multivariate linear models, but have not yet been developed for constrained models that we develop in this article.

We first compare the standard envelope estimator based on an unconstrained multivariate  model with the standard estimator arising from a constrained multivariate model in terms of bias and efficiency. Then, to further improve efficiency, we propose a novel envelope estimator based on a constrained multivariate model.  Novel envelope-based testing methods are also proposed. We provide support for our proposals  by simulations and by studying the classical dental data and  data from the China Health and Nutrition Survey {\color{black}{and a study of probiotic capacity to reduced Salmonella infection }}.}

\end{abstract}

\section{Introduction}\label{sec:intro}
Consider the multivariate linear regression model
\begin{eqnarray}\label{mlm}
\Y_{i} = \betabf_{0} + \betabf \X_{i} + \varepsilonbf_{i},  \; i=1,\ldots,n,
\end{eqnarray}
where the stochastic response $\Y_{i} \in \real{r}$, the non-stochastic  predictor vectors $\X_{i} \in \real{p}$, $\betabf_{0} \in \real{r}$, 
and the error vectors $\varepsilonbf_{i}$ are independent copies  of $\varepsilonbf \sim N(0, \Sigmabf)$. Model (\ref{mlm}) is {\em unconstrained} in the sense that each response is allowed a separate linear regression:  the maximum likelihood estimator of the $j$-th row of $(\betabf_{0},\betabf)$ is the same as the estimator of the coefficients from the linear regression of the $j$-th response on  $\X$. 

 {In many applications, particularly analyses of growth curves and longitudinal data,  there may be additional information that $\spn(\betabf_{0},\betabf)$ is  contained in a {\em known} subspace $\uspc$ with basis matrix $\U \in \real{r \times k}$.   The classic dental data \citep{PotthoffRoy1964,Lee1975,Rao1987,Lee1988} provides an illustration of this type of structure.}

\begin{example}\label{eg: dental_data}
A study of dental growth measurements of the
distance (mm) from the center of the pituitary gland to the
pteryomaxillary fissure were obtained on 11 girls and 16 boys at ages 8, 10, 12, and 14. The goal was to study the growth measurement as a function of time and  sex.
\end{example}
Let $Y_{ik}$ denote the continuous measure of distance for child $i$ at age $t_{k}$, for $t_{k}=8$, $10$, $12$, $14$, and let $\X_{i}$ denote the gender indicator for child $i$ (1 for boy and 0 for girl). After graphical inspection, many researchers treated the population means for distance as linear in time for each gender.  Following this tradition, a mixed effects repeated
 measure model is
$Y_{ik}=\alpha_{00}+b_{0i}+\alpha_{01}\X_i+(\alpha_{10}+b_{1i}+\alpha_{11}\X_{i})t_{k}+\varepsilon_{ik}^{*},$ 
 where $\varepsilonbf_{i}^{*}=(\varepsilon_{i1}^{*},\dots,\varepsilon_{ir}^{*})^T\overset{\text{i.i.d.}}{\sim}N(\zerobf,\Sigmabf^*)$, $b_{0i}$ and $b_{1i}$ denote the random intercept and slope, $(b_{0i},b_{1i})\overset{\text{i.i.d.}}{\sim}N(\zerobf,\D)$, where$\D$ is a $2 \times 2$ positive difinite matrix. 
We rewrite this model  as 
\begin{eqnarray}\label{gmlm} \Y_{i}& =  &\U \alphabf_0 + \U \alphabf \X_i + \epsilon_i
\end{eqnarray}
with $\U := (\mathbf{1},\mathbf{t})$ with $\mathbf{t}=(8,10,12,14)^T$, $\alphabf_0= (\alpha_{00}, \alpha_{10})$
$\alphabf=(\alpha_{01},\alpha_{11})$, $\varepsilonbf_i=\varepsilonbf_i^{*}+b_{0i}\onebf_{r\times1}+b_{1i}\mathbf{t}$ and $\varepsilonbf_i\overset{\text{i.i.d.}}{\sim}N(\zerobf,\Sigmabf)$. 
%
Applying the same ideas to just $\betabf$, so $\spn(\betabf) \subseteq \uspc$ without requiring that $\spn(\betabf_{0}) \subseteq \uspc$,  leads to the model
\begin{eqnarray}\label{igmlm}
\Y_{i} = \betabf_{0} + \U \alphabf \X_{i} + \varepsilonbf_{i},  \; i=1,\ldots,n.
\end{eqnarray}
Let  $\bspc = \spn(\betabf)$.  
If we set $\U = \1_{r}$, so in model (\ref{igmlm}) $\alphabf$ is a row vector of length $p$, then the mean functions for the individual responses are parallel.  {Although motivated in the context of the dental data, we use models (\ref{gmlm}) and (\ref{igmlm}) as general forms that can be adapted for different applications by choice of $\U$, referring to them as {\em constrained} multivariate linear models. }   \citet{Cooper2002} used a version of model (\ref{gmlm}) with $\U$ reflecting charge balance constraints on chemical constituents of water samples.  

 Constrained models occur in various areas including growth curve and longitudinal studies where the elements of $\Y_{i}$ are repeated observations on the $i$-th experimental unit over time.  It is common in such settings to model the rows of $\U$ as a user-specified vector-valued function $\bu(t)\in \real{k}$ of time $t$, the $i$-th row of $\U$ then being $\bu^{T}(t_{i})$.  Polynomial bases $\bu^{T}(t) = (1, t, t^{2},\ldots,t^{k-1})$ are prevalent, particularly in the foundational work of \citet{PotthoffRoy1964}, \citet{Rao1965}, \citet{Grizzle1969} and others, but splines \citep{Nummi2008} or other basis constructions \citep{Izenman1989} could be used as well. In longitudinal studies, model (\ref{gmlm}) might be used when it is desirable to model  profiles, while model (\ref{igmlm}) could be used when modeling just profile differences.  For instance, if $\X = 0,1$ is a population indicator then under model (\ref{gmlm}) the mean profiles are modeled as $\U\alphabf_{0}$ and $\U(\alphabf_{0} + \alphabf)$, while under model (\ref{igmlm}) the profile means are $\betabf_{0}$ and $\betabf_{0} + \U\alphabf$. It is known in the literature that constrained models gain efficiency in the estimators compare with model (\ref{mlm}), provided that $\U$ is correctly specified.
However, it may be very difficult to correctly specify $\U$ in some applications, as in the following study from  \cite{Kenward1987}.
\begin{example}\label{otro}
An experiment was carried out to compare two treatments for the control of gut worm in cattle.  The treatments were each randomly assigned to 30 cows whose weights were measured at $2$, $4$, $6,\ldots,18$ and $19$ weeks after treatment.  The goal of the experiment was to see if a differential treatment effect could be  detected and, if so, the time point when the difference was  first manifested.  
\end{example}
\begin{figure}[ht!]
    \centering
        \includegraphics[scale=.5]{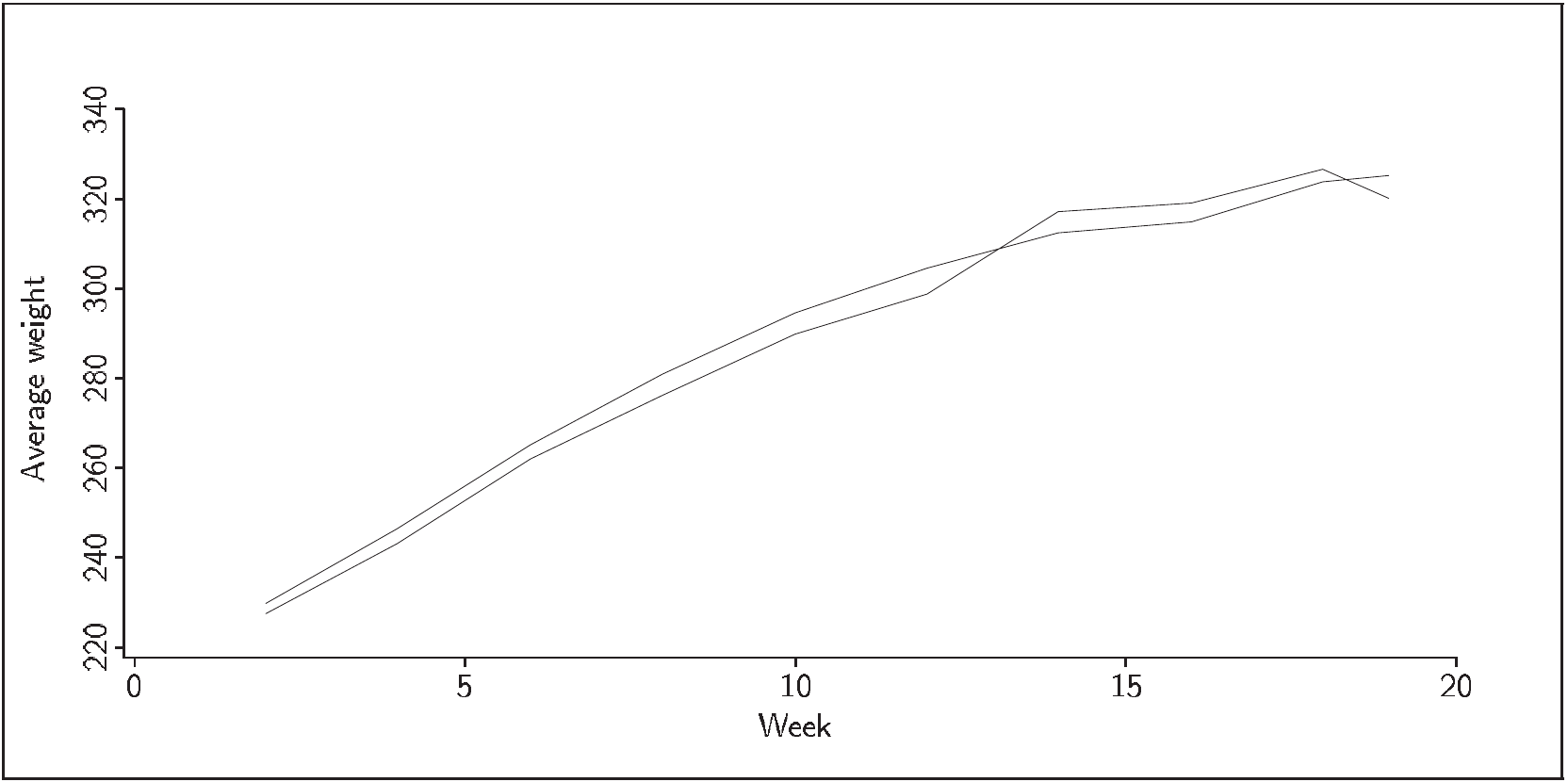}
        \caption{Cattle data: Average weight by treatment and time. }
        \label{cows}
\end{figure}
The constrained models (\ref{gmlm}) and (\ref{igmlm}) require that we select $\uspc$.  Lacking prior knowledge, it is natural to inspect plots of the average weight by time, as shown in Figure~\ref{cows}.  It seems clear from the figure that it would be difficult to model the treatment profiles, particularly their two crossing points, without running into problems of over fitting.  {Envelopes provided a way to model data like that illustrated in Figure~\ref{cows} without specifying a subspace $\uspc$.}


 {Envelope methodology is based on a relatively new paradigm for dimension reduction that, when applied in the context of model (\ref{mlm}),  has some similarity with constrained multivariate models.   Briefly, envelopes produce a re-parameterization of model (\ref{mlm}) in terms of a basis $\Gammabf \in {\mathbb R}^{r\times u}$ for the smallest reducing subspace of $\Sigmabf$ that contains $\bspc$.  Like the constrained model, envelopes produce an upper bound for $\bspc$, $\bspc \subseteq \spn(\Gammabf)$, but unlike the constrained model the bound is unknown and must be estimated.  Also, unlike the constrained model, $\Gammabf^{T}\Y$ contains the totality of $\Y$ that is affected by changing $\X$.  Since $\bspc \subseteq \spn(\Gammabf)$, we have $\betabf = \Gammabf \etabf $ for some $\etabf \in \real{u \times p}$.  Model (\ref{mlm}) can be then be re-paremeterized to give its envelope counterpart,
\begin{eqnarray}\label{envmlm}
\Y_{i} & = & \betabf_{0} +\Gammabf \etabf \X_{i} + \varepsilonbf_{i},  \; i=1,\ldots,n, \\
\Sigmabf & = & \Gammabf \Omegabf \Gammabf^{T} +  \Gammabf_{0} \Omegabf_{0}\Gammabf_{0}^{T}, \nonumber
\end{eqnarray}
where   $(\Gammabf, \Gammabf_{0})\in \real{r \times r}$,  orthogonal, $\Omegabf = \Gammabf^{T}\Sigmabf\Gammabf > 0$ and $\Omegabf _{0}= \Gammabf_{0}^{T}\Sigmabf\Gammabf_{0} > 0$. Envelopes are reviewed in more detail in Section \ref{sec:review}.}

Comparing (\ref{gmlm})--(\ref{igmlm}) with (\ref{envmlm}), both express $\betabf$ as a basis times a coordinate matrix: $\betabf = \U\alphabf$ in (\ref{gmlm})--(\ref{igmlm}) and $\betabf = \Gammabf \etabf$ in (\ref{envmlm}). However, as mentioned previously, $\Gammabf$ is estimated but $\U$ is assumed known.
Envelopes were  first proposed by \citet*{Cook2007} to facilitate dimension reduction and later were shown by \cite{Cook2010} to have the potential to achieve massive efficiency gains relative to the standard maximum likelihood estimator of $\betabf$, and that these gains will be passed on to other tasks such as prediction. There are now a number of extensions and applications of this basic envelope methodology, each demonstrating the potential for substantial efficiency gains \citep{SuCook2011,CookZhang2015foundation,CookZhang2015simultaneous,ForzaniSu,Su2016,LiZhang2017tensor,Hossien2017}.  
Studies over the past several years have  demonstrated repeatedly that sometimes the efficiency gains of the envelope methods relative to standard methods amount  to increasing the sample size many times over.
 See \citet{CookEnv2018} for a review and additional extensions of envelope methodology.




{The choice between a constrained model, (\ref{gmlm}) or (\ref{igmlm}), and the envelope model (\ref{envmlm}) hinges on the ability to correctly specify an upper bound $\uspc$ for $\spn(\betabf_{0},\betabf)$ or $\bspc$. As we show  in Section \ref{sec:comparison}, if we have a correct parsimonious basis $\U$
 then the constrained models are more efficient.
But if bias is present or if we use a correct but excessive $\U$, then the envelope model (\ref{envmlm})  can be much more efficient. 
Although considerable methodology has been developed for the envelope version (\ref{envmlm}) of the unconstrained model (\ref{mlm}), there are apparently no envelope counterparts available for the class of models represented by (\ref{gmlm}) and (\ref{igmlm}) when a correct parsimonious $\U$ is available. In Section 
\ref{sec:Balpha} we show how to adapt the envelope paradigm to models  (\ref{gmlm}) and (\ref{igmlm}) to achieve efficiency gains over those models.  Testing methods are proposed 
in Section \ref{sec:testing} to evaluate the choice of $\U$ and also to test importance of predictors.  Simulations to support our finding are given in Section~\ref{simulations} and  in Section~\ref{sec:applications} we compare our methodology with others in { \color{black}{two}} examples. Proofs for all propositions and discussions of related issues are available in a Supplement to this article.
} 


\paragraph{Notational conventions.} 
 Given a sample $(\abf_{i}, \bb_{i}), i=1,\ldots,n$, let $\T_{\abf,\bb} = n^{-1}\sum_{i=1}^{n} \abf_{i}\bb_{i}^{T}$ denote the matrix of raw second moments, and let $\T_{\abf} = n^{-1}\sum_{i=1}^{n} \abf_{i}\abf_{i}^{T}$.  For raw second moments involving $\Y_{S}$ and $\Y_{D}$ (defined herein) we use $S$ and $D$ as subscripts.  We use a subscript 1 in residuals computed from a model containing a vector of intercepts.  The absence of a 1 indicates no intercept was included.  For instance, $\R_{\abf | \bb}$ means the residuals from the regression of $\abf$ on $\bb$ without an intercept vector, $\abf_{i} = \betabf \bb_{i} + \e$, while $\R_{\abf | (1, \bb)}$ means the residuals from the regression of $\abf$ on $\bb$ with an intercept vector, $\abf_{i} = \betabf_{0} + \betabf \bb_{i} + \e_{i}$. Similarly, $\R_{D|S}$ means a residual from the regression of $\Y_{D}$ and $\Y_{S}$ without an intercept, while $\R_{D|(1,S)}$  means a residual from the regression of $\Y_{D}$ and $\Y_{S}$ with an intercept.

Sample variances are written as $\Sbf_{\abf} = n^{-1}\sum_{i=1}^{n}(\abf_{i} - \bar{\abf})(\abf_{i} - \bar{\abf})^{T}$ and sample covariance matrices are written as $\Sbf_{\abf, \bb} = n^{-1}\sum_{i=1}^{n}(\abf_{i} - \bar{\abf})(\bb_{i} - \bar{\bb})^{T}$.  For variances and covariances involving $\Y_{D}$ and $\Y_{S}$ we again use $D$ and $S$ as subscripts, e.g. $\Sbf_{D} = n^{-1}\sum_{i=1}^{n}(\Y_{Di} - \bar{\Y}_{D})(\Y_{Di} - \bar{\Y}_{D})^{T}$.  The notation $\Sbf_{\abf|\bb}$ means the covariance matrix of the residuals from fit of the model $\abf_{i} = \betabf_{0} + \betabf \bb_{i} + \e_{i}$, which always includes as intercept.  That is, $\Sbf_{\abf|\bb} = n^{-1}\sum_{i=1}^{n}\R_{\abf | (1, \bb), i}\R_{\abf | (1, \bb), i}^{T}$.  Similarly, $\Sbf_{D|S} = \sum_{i=1}^{n}\R_{D | (1, S), i}\R_{D | (1, S), i}^{T}$.

We use $\spn(\A)$ to denote the subspace spanned by the columns of the matrix $\A$.  The projection onto  $\spc = \spn(\A)$ will be denoted using either the subspace itself  $\Pbf_{\spc}$ or its basis  $\Pbf_{\A}$.   Projections onto an orthogonal complement will be denoted similarly using $\Q_{(\cdot)} = \I-\Pbf_{(\cdot)}$. For a subspace $\spc$ and conformable matrix $\B$, $\B \spc = \{\B S \mid S \in \spc\}$. If an estimator $\ba \in \real{r}$ of $\alphabf \in \real{r}$  has the property that $\sqrt{n}(\ba - \alphabf)$ is asymptotically normal with mean 0 and variance $\A$, we write  $\avar(\sqrt{n}\ba)=\A$ to 
denote  its asymptotic variance.

\section{Comparison of the envelope and constrained estimators 
}\label{sec:comparison}

Models  (\ref{gmlm})--(\ref{igmlm}) and (\ref{envmlm}) are similar in the sense that $\betabf$ is represented as a basis  times a coordinate matrix, $\betabf = \U\alphabf$ in (\ref{gmlm})--(\ref{igmlm}) and $\betabf = \Gammabf \etabf$ in (\ref{envmlm}).  It might be thought that (\ref{gmlm}) and (\ref{igmlm})  would yield the better estimators because $\U$ is known while $\Gammabf$ is not, but that turns out not to be so generally, in part because we may have $\bspc \not \subseteq \uspc$, which raises the issue of bias as discussed in Section~\ref{sec:bias}, and in part because the envelope model capitalizes automatically on structure in $\Sigmabf$, which can improve efficiency as discussed in Section~\ref{sec:varcomparison}.  Our general conclusion is that in practice it may be necessary to compare their fits before selecting an estimator, and that the envelope estimator may have a clear advantage when there is uncertainty in the choice of $\uspc$, as illustrated in Figure~\ref{cows}.

 {Developments under models (\ref{gmlm}) and (\ref{igmlm}) are very similar since they differ only on how the intercept is handled.  In the remainder of this article we focus on model (\ref{gmlm}) and comment from time to time on modifications necessary for model (\ref{igmlm}). All subsequent developments are under model (\ref{gmlm}) unless model (\ref{igmlm}) is indicated explicitly.}
 
 \subsection{Maximum likelihood estimators for constrained models}\label{sec:gmlm}

Our treatment of maximum likelihood estimation from (\ref{gmlm})  is based on  linearly  transforming $\Y$.
Let $\U_{0}$ be a semi-orthogonal basis matrix for $\uspc^{\perp}$, and let $\W=(\U(\U^{T}\U)^{-1}, \U_{0}) := (\W_{1},\W_{2})$.  Then the transformed model becomes
    \begin{equation}\label{tmodel}
\W^{T}\Y_{i} :=  \left(\begin{array}{c} \Y_{Di} \\ \Y_{Si} \end{array}\right)=\left(\begin{array}{c} (\U^{T}\U)^{-1}\U^{T}\Y_{i} \\ \U_{0}^{T}\Y_{i}\end{array}\right)=\left(\begin{array}{c} \alphabf_{0}+\alphabf\X_{i} \\ 0 \end{array}\right) + \W^{T}\varepsilonbf_{i},\; i=1,\ldots,n,
\end{equation}
{ where $\Y_{Di} \in \real{k}$ and $\Y_{Si}\in \real{r-k}$} with $k$  the number of columns of $\U$. The transformed variance can be represented block-wise as $\Sigmabf_{\W} := \var(\W^{T}\varepsilonbf) = (\W_{i}^{T}\Sigmabf\W_{j})$, $i,j=1,2$, where $\Sigmabf$ is as defined for model (\ref{gmlm}).
 The mean $E(\Y_{D} \mid \X)$ depends non-trivially on $\X$ and thus, as indicated by the subscript $D$, we think of $\Y_{D}$ as providing direct information about the regression.  On the other hand, $E(\Y_{S}\mid \X)=0$ and thus $\Y_{S}$ provides no direct information but may provide useful subordinate information by virtue of its association with $\Y_{D}$.  

To find the maximum likelihood estimators from model (\ref{tmodel}), we write the full log likelihood as the sum of the log likelihoods for the marginal model for $\Y_{S}\mid \X$ and the conditional model for $\Y_{D} \mid (\X,\Y_{S})$:
\begin{eqnarray}
\Y_{Si} \mid \X& = &  \e_{Si} \label{cmodelS}\\
\Y_{Di}\mid(\X_{i}, \Y_{Si})  & = & \alphabf_0 + \alphabf \X_{i} + \phibf_{D|S}\Y_{Si}  + \e_{D|Si}, \label{cmodelD}
\end{eqnarray}
where $\phibf_{D|S} = (\U^{T}\U)^{-1}\U^{T}\Sigmabf \U_{0}(\U_{0}^{T}\Sigmabf\U_{0})^{-1} \in \real{k \times (r-k)}$, $\e_{D|S} = \W_{1}^{T}\varepsilonbf$, $\e_{S} = \W_{2}^{T}\varepsilonbf$.  The variances of the errors are 
$\Sigmabf_{S}  :=\var(\e_{S}) =\U_0^T \Sigmabf \U_0 $ and 
$\Sigmabf_{D|S}   :=   \var(\e_{D|S})= ( \U^{T}\Sigmabf^{-1}\U)^{-1}$.
 The number of free real parameters in this conditional model is $N_{\Cm}(k)=k(p+1) + r(r+1)/2$. The subscript `$\Cm$' is used also to indicate estimators arising from the conditional model (\ref{cmodelD}).  The maximum likelihood estimator and its   asymptotic variance are
 \begin{eqnarray}
 \betabfhat_{\Cm} &=& \U \alphabfhat_{\Cm} =  \U \Sbf_{D,\R_{\X|(1,S)}} \Sbf_{\X|S}^{-1} = \U (\Sbf_{D,\X} - \Sbf_{D,S} \Sbf_{S}^{-1}\Sbf_{S,\X})\Sbf_{\X|S}^{-1}
 \label{alphahatcm}
 \\
\avar(\sqrt{n}\vecc(\betabfhat_{\Cm})) 
& = &  \Sigmabf_{\X}^{-1}\otimes \U \Sigmabf_{D\mid S}\U^{T}, \label{avarbetahatcm}
\end{eqnarray}
where $\alphabfhat_{\Cm}$ and $\betabfhat_{\Cm}$ are the MLEs of $\alphabf$ and $\betabf$ in model (\ref{tmodel}). 

Estimation for model (\ref{igmlm}) requires just a few modifications of the procedure for model (\ref{gmlm}).  All modifications stem from the presence of an intercept vector in model (\ref{cmodelS}), which becomes $\Y_{S} = \W_{2}^{T}\betabf_{0} + \e_{S}$.  In consequence, the variance $\Sigmabf_{S}$ is estimated as $\Sigmabfhat_{S} = \Sbf_{S}$ with corresponding change in the estimator of $\Sigmabf$, and the estimator of the intercept $\W_{2}^{T}\betabf_{0}$ is just $\Ybar_{S}$. The intercept in (\ref{cmodelD}) is redefined as $\alphabf_{0} = \W_{1}^{T}\betabf_{0}-\phibf_{D|S}\W_{2}^{T}\betabf_{0}$.  The maximum likelihood estimator of $\betabf_{0}$ in model (\ref{igmlm}) can be constructed straightforwardly from the estimators of $\alphabf_{0}$, $\W_{2}^{T}\betabf_{0} $ and $\phibf_{D|S}$.  Because there is an intercept in (\ref{cmodelS}), the number of real parameters becomes $N_{\Cm} + r-k$.  Importantly, the estimators of the parameters  in (\ref{cmodelD}) are unchanged.  This means in particular that $\alphabfhat_{\Cm}$ and $\betabfhat_{\Cm}$ along with their  asymptotic variances are the same under models (\ref{gmlm}) and (\ref{igmlm}), although different $\U$'s might be used in their construction.

\subsection{Envelope estimator stemming from  Model \eqref{mlm}}\label{sec:review}


Consider a subspace $\spc \subseteq \real{r}$ that satisfies the two conditions (i) $\Q_{\spc}\Y\mid \X=\x_{1} \sim \Q_{\spc}\Y\mid \X=\x_{2}$ for all relevant $\x_{1}$ and $\x_{2}$ and (ii) $\Pbf_{\spc}\Y \indep \Q_{\spc}\Y \mid \X$.   Condition (i) insures that the marginal distribution of $\Q_{S}\Y$ does not depend on $\X$, while statement (ii) insures that, given $\X$, $\Q_{\spc}\Y$ cannot provide material  information via an association with $\Pbf_{\spc}\Y$.  Together these conditions imply that the impact of $\X$ on the distribution of $\Y$ is concentrated solely in $\Pbf_{\spc}\Y$.  One motivation underlying envelopes is then to characterize linear combinations $\Q_{S}\Y$ that are unaffected by changes in $\X$ and in doing so produce downstream gains in estimative and predictive efficiency.  

  In terms of model (\ref{mlm}), condition (i) holds if and only if $\bspc\subseteq \spc$ and condition (ii) holds if and only if $\spc$ is a reducing subspace of $\Sigmabf$; that is, $\spc$ must decompose  $\Sigmabf = \Pbf_{\spc} \Sigmabf \Pbf_{\spc} + \Q_{\spc}\Sigmabf \Q_{\spc}$. The intersection of all subspaces with these properties is by construction the smallest reducing subspace of $\Sigmabf$ that contains $\bspc$, which is called the $\Sigmabf$-envelope of $\bspc$ and is represented as $\espc_{\Sigmabfs}(\bspc)$ \citep{Cook2010}.  These consequences of conditions (i) and (ii) can be incorporated into model (\ref{mlm}) by using a basis, leading to model \eqref{envmlm}.  Let $u \in \{0,1,\ldots,r\}$ denote the  dimension of $\espc_{\Sigmabfs}(\bspc)$. The number of free real parameters is $N_{\Em} = r + pu + r(r+1)/2$. The subscript `$\Em$' is used also to indicate selected quantities arising from this envelope model.  The goal here still to estimate $\betabf = \Gammabf\etabf$ and $\Sigmabf$.  
 \citet{Cook2010} derived the maximum likelihood envelope estimators of $\betabf$ and $\Sigmabf$ along with their asymptotic variances. They showed that substantial efficiency gains in estimation of $\betabf$ are possible under this model, particularly when a norm of $\var(\Gammabf_{0}^{T}\Y) =\Omegabf_{0}$ is considerably larger than the same norm of $\var(\Gammabf^{T}\Y)=\Omegabf$. 

Given the envelope dimension $u$, the maximum likelhood estimator $\betabfhat_{\Em}$ of $\betabf = \Gammabf \etabf $ from envelope model (\ref{envmlm}) has asymptotic variance given by 
\begin{equation}\label{avarbetahatem}
\avar(\sqrt{n}\vecc(\betabfhat_{\Em})) =  \Sigmabf_{\X}^{-1}\otimes \Gammabf \Omegabf \Gammabf^T + (\etabf^T\otimes \Gammabf_0) \M^{\dagger}(\Sigmabf_{\X}) (\etabf \otimes \Gammabf^T_0),
\end{equation}
where for a $\C \in \real{p \times p}$, $\M (\C) :=   \etabf  \C\etabf^T \otimes \Omegabf_0^{-1} + \Omegabf \otimes \Omegabf_0^{-1} + \Omegabf^{-1} \otimes \Omegabf_0 - 2\I$ and $\dagger$ denotes the Moore-Penrose inverse. \cite{Cook2010} showed that $\avar(\sqrt{n}\vecc(\betabfhat_{\Em})) \leq \avar(\sqrt{n}\vecc(\betabfhat_{\Um}))$, where $\betabfhat_{\Um}$ is the maximum likelihood estimator under the uncontrained model \eqref{mlm}.  In consequence, estimators from the envelope model (\ref{envmlm}) are always superior to those from the unconstrained multivariate model (\ref{mlm}). \cite{Cook2010} also showed that the envelope estimator is {  $\sqrt{n}$-consistent} even the normality assumption is violated as long as the data has finite fourth moments.

\subsection{Potential bias in $\betabfhat_{\Cm}$}\label{sec:bias}

Assuming that $\bspc \subseteq \uspc$, $\alphabfhat_{\Cm}$ and $\betabfhat_{\Cm}$ are unbiased estimators of $\alphabf$ and $\betabf$.  However, if $\bspc \not\subseteq \uspc$ then both $\alphabfhat_{\Cm}$ and $\betabfhat_{\Cm}$ are biased, which could materially affect the estimators: $E(\alphabfhat_{\Cm}) = (\U^{T}\U)^{-1}\U^{T}\betabf$ and $E(\betabfhat_{\Cm}) = \Pbf_{\U}\betabf$.  Consequently, the bias in $\betabfhat_{\Cm}$ is $\betabf -  \Pbf_{\U}\betabf = \Q_{\U}\betabf$.  A nonzero bias must necessarily dominate the mean squared error asymptotically and so could limit the utility of $\betabfhat_{\Cm}$. Simulation results that illustrate the potential bias effects are discussed in Section \ref{sec: sim_bias}.  Otherwise, we assume that $\bspc \subseteq \uspc$ in the remainder of this article unless indicated otherwise.

\subsection{Comparison of asymptotic variances of $\betabfhat_{\Em}$ and $\betabfhat_{\Cm}$}\label{sec:varcomparison}

We now compare the asymptotic variance  of the envelope and constrained estimators of $\betabf$, (\ref{avarbetahatem}) and (\ref{avarbetahatcm}).  Depending on the dimensions involved, the relationship between $\uspc$ and the envelope $\espc_{\Sigmabfs}(\bspc)$ and other factors, the difference between the asymptotic covariance matrices for the estimators -- $\betabfhat_{\Em}$ and $\betabfhat_{\Cm}$ -- from these two models can be positive definite, negative definite or indefinite. 
 Since all comparisons are in terms of $\betabf$'s, we assume without loss of generality that $\U$ is a semi-orthogonal matrix.  Also, since $\betabfhat_{\Cm}$ is the same under models (\ref{gmlm}) and (\ref{igmlm}) we do not distinguish between these models in this section.

\subsubsection{$\bspc \subseteq \uspc \subseteq \espc_{\Sigmabfs}(\bspc)$}  

Assuming that $\uspc$ is correct so $\bspc \subseteq \uspc $ and that $\uspc \subseteq \espc_{\Sigmabfs}(\bspc)$ is one way to simplify the variance comparison: 

\begin{prop}\label{lemmacmltem}
If  $\bspc \subseteq \uspc \subseteq \espc_{\Sigmabfs}(\bspc)$,  then $\avar(\sqrt{n}\vecc(\betabfhat_{\Cm})) \leq \avar(\sqrt{n}\vecc(\betabfhat_{\Em})).$
\end{prop}
In consequence, under the hypothesis  $\bspc \subseteq \uspc \subseteq \espc_{\Sigmabfs}(\bspc)$, the constrained estimator $\betabfhat_{\Cm}$ is superior to the envelope estimator $\betabfhat_{\Em}$.  However, this  comparison may be seen as loaded in favor of  $\betabfhat_{\Cm}$ since the constrained estimator uses the additional knowledge that $\bspc \subseteq \uspc$, the envelope estimator  does not.  Additionally, neither estimator makes use of the proposition's hypothesis.   The next proposition provides help in assessing the impact of the hypothesis on the underlying structure by connecting it with $\espc_{\Sigmabfs}(\uspc)$, the $\Sigmabf$-envelope of $\uspc$.

\begin{prop}\label{lemma3} $\mathrm{\;}$ Assume that $\bspc \subseteq \uspc$.  Then
\begin{enumerate}
\item  $\espc_{\Sigmabfs}(\bspc) \subseteq \espc_{\Sigmabfs}(\uspc)$,
\item $\uspc \subseteq  \espc_{\Sigmabfs}(\bspc)$
if and only if $\espc_{\Sigmabfs}(\bspc) = \espc_{\Sigmabfs}(\uspc)$, 
\item If $\rank(\alphabf) = k$ then $\bspc = \uspc$ and $\espc_{\Sigmabfs}(\bspc) = \espc_{\Sigmabfs}(\uspc)$.
\end{enumerate}
 \end{prop}
This proposition says essentially that if $\bspc \subseteq \uspc \subseteq  \espc_{\Sigmabfs}(\bspc)$ then we can start with model (\ref{mlm}) and parameterize in terms of $\espc_{\Sigmabfs}(\uspc)$ rather than $\espc_{\Sigmabfs}(\bspc)$. A key distinction here is that $\uspc$ is known while $\bspc$ is not.  In consequence, we expect less estimative variation when parameterizing (\ref{mlm}) in terms of $\espc_{\Sigmabfs}(\uspc)$ instead of $\espc_{\Sigmabfs}(\bspc)$.  Since $\uspc \subseteq \espc_{\Sigmabfs}(\uspc)$ can construct a  semi-orthogonal basis  for $\espc_{\Sigmabfs}(\uspc)$ as $\Gammabf = (\U, \Gammabf_{2})$ with $\U_{0} = (\Gammabf_{2},\Gammabf_{0})$ and, recognizing that $\betabf = \U\alphabf = \Gammabf\etabf$, we get a new  model
 \begin{eqnarray}\label{gmlmB}
\Y_{i} & = & \U \alphabf_{0} + \U \alphabf \X_{i} + \varepsilonbf_{i},  \; i=1,\ldots,n \\
\Sigmabf & = &  \Gammabf \Omegabf \Gammabf^{T} + \Gammabf_{0}\Omegabf_{0}\Gammabf_{0}^{T}.\nonumber
\end{eqnarray}
Consider estimating $\alphabf$ from this model using the steps sketched in Section~\ref{sec:gmlm}, and partition $\Omegabf = (\Omegabf_{ij})$ to conform to the partition of  $\Gammabf = (\U, \Gammabf_{2})$.  The envelope structure of (\ref{gmlmB}) induces a special structure on the reduced model that corresponds to (\ref{cmodelS})--(\ref{cmodelD}): $\Sigmabf_{S} = \mathrm{bdiag}(\Omegabf_{22}, \Omegabf_{0})$ is block diagonal, $\Sigmabf_{D|S} = \Omegabf_{11} - \Omegabf_{12}\Omegabf_{22}^{-1}\Omegabf_{21}$ and $\phibf_{D|S} = (\Omegabf_{12}\Omegabf_{22}^{-1}, 0)$.   
It can now be shown that the estimators of $\alphabf$ from the constrained model (\ref{cmodelS})--(\ref{cmodelD}) and from (\ref{gmlmB}) have the same asymptotic variance.  In other words, if we neglect the hypothesized condition that $\uspc \subseteq \espc_{\Sigmabfs}(\bspc)$ then the constrained estimator is better, but if we formulate the envelope model making use of that condition then the constrained and envelope estimators are asymptotically equivalent.


\citet{Rao1967} posited a simple structure for the analysis of balanced growth curve data \citep[See also][]{Geisser1970, Lee1975, Geisser1981, Lee1988, PanFang2002}.   In our context, Rao's simple structure is obtained by assuming that $\espc_{\Sigmabfs}(\uspc) = \uspc$, which corresponds to model (\ref{gmlmB}) with $\Gammabf = \U$.  In view of the options available, Rao's simple structure seems too specialized to warrant further attention. Additional discussion of Rao's simple structure is available in Supplement Section~\ref{sec:structure}.

\subsubsection{$\uspc \supseteq  \espc_{\Sigmabfs}(\bspc)$} 

Assuming that $\uspc \supseteq \espc_{\Sigmabfs}(\bspc)$ is another way to simplify the variance comparison.  Let  $\Gammabf \in \real{r \times u}$ be a semi-orthogonal basis matrix for $\espc_{\Sigmabfs}(\bspc)$ and let $(\Gammabf, \Gammabf_{0})$ be an orthogonal matrix. Since $\uspc \supseteq  \espc_{\Sigmabfs}(\bspc)$, we can construct semi-orthogonal bases $\U = (\Gammabf, \Gammabf_{01})$ and $\Gammabf_{0}  = (\Gammabf_{01}, \Gammabf_{02})$. Partition $\Omegabf_{0} = (\Omegabf_{0, ij})$ to correspond to the partitioning of $\Gammabf_{0}$.  Then
\begin{prop}\label{lemmaavar}
Assume that $\uspc \supseteq  \espc_{\Sigmabfs}(\bspc)$ and let $\cbf \in \real{r}$. Then 
\begin{enumerate}
\item If $\cbf \in \espc_{\Sigmabfs}(\bspc)$ then $\avar(\sqrt{n}\cbf^{T}\betabfhat_{\Cm}) = \avar(\sqrt{n}\cbf^{T}\betabfhat_{\Em})$.
\item If $\cbf \in \spn(\Gammabf_{02})$ then $\avar(\sqrt{n}\cbf^{T}\betabfhat_{\Cm}) \leq \avar(\sqrt{n}\cbf^{T}\betabfhat_{\Em})$.
\item  If $\cbf \in \spn(\Gammabf_{01})$, $\rank( \M(\Sigmabf_{\X})) = \rank(\etabf\Sigmabf_{\X}\etabf^{T} \otimes \Omegabf_{0}^{-1})$ and $\Omegabf_{12} = 0$ then $\avar(\sqrt{n}\cbf^{T}\betabfhat_{\Cm}) \geq \avar(\sqrt{n}\cbf^{T}\betabfhat_{\Em})$.
\end{enumerate}
\end{prop}
The central message of this lemma is  that the difference between the asymptotic covariance matrices for the estimators $\betabfhat_{\Em}$ and $\betabfhat_{\Cm}$ can be positive semi-definite or negative semi-definite, depending on the characteristics of problem.
 
Although the above derivation is under two simple cases where $\mathcal{U}$ and the envelope space
are nested, the conclusion actually holds for the general case: {if we have a correct parsimoniously parameterized constrained model then the envelope model (\ref{envmlm}) is less efficient;
but if the basis $\U$ in the constrained model is incorrect or if the constrained model is excessively parameterized, then envelopes can be much more efficient.} This motivated us to incorporate envelopes into the constrained model so that we can further improve efficiency if constraints are reasonably well modeled for the data

 \section{Envelopes  in constrained models} \label{sec:Balpha}
 
{In this section, we consider two different ways of imposing envelopes in a constrained model when $\bspc \subseteq \mathcal{U}$. As mentioned previously, we focus on envelope estimators in the constrained model (\ref{gmlm}) and later describe the modifications necessary for model (\ref{igmlm}). 
 In Section~\ref{sec:envalpha} we describe envelope estimation of $\alphabf$ when there is available an application-grounded basis $\U$ that is key to interpretation and inference.  In Section~\ref{sec:envbeta} we address envelope estimation of $\betabf = \U \alphabf$.  Here the choice of basis $\U$ has no effect on the maximum likelihood estimators of $\betabf$ under the constrained models (\ref{gmlm}),  but it does affect the envelope estimator of $\betabf$.  Basis selection is addressed in Section~\ref{sec:envbeta}.}

\subsection{Enveloping $\alphabf$}\label{sec:envalpha}

{ Estimation of $\alphabf$ will be of interest when it is desirable to interpret $\betabf = \U \alphabf$ in terms of its coordinates $\alphabf$ relative to the known application-grounded basis $\U$.  Let $\aspc = \spn(\alpha)$. The envelope estimator of $\alphabf$ in model (\ref{tmodel}) can be found by first transforming  \eqref{tmodel} into \eqref{cmodelS}--\eqref{cmodelD} and then parameterizing (\ref{cmodelD}) in terms of a semi-orthogonal basis matrix $\Phibf \in \real{k \times u}$ for $\espc_{\Sigmabfs_{D|S}}(\aspc)$, the $\Sigmabf_{D|S}$-envelope of $\aspc$ with dimension  $u \leq k$.  Since $\avar(\sqrt{n}\vecc(\alphabfhat_{\Cm})) = \Sigmabf_{\X}^{-1} \otimes \Sigmabf_{D|S}$ is in the form of a Kronecker product that allows separation of row and column effects of $\alphabf$, this structure follows also from the theory of \citet{CookZhang2015foundation, CookZhang2015simultaneous} for matrix-valued envelope  estimators based on envelopes of the form $\real{p}\oplus \espc_{\Sigmabfs_{D|S}}(\aspc)$, where $\oplus$ denotes the direct sum. }

 Let $\etabf \in \real{u \times p}$ be an unconstrained matrix giving the coordinates of $\alphabf$ in terms of semi-orthogonal basis matrix $\Phibf$, so $\alphabf = \Phibf \etabf$, and let $(\Phibf, \Phibf_{0}) \in \real{k \times k}$ be an orthogonal matrix.  Then the envelope version of model (\ref{cmodelS})--(\ref{cmodelD}) is a version of the partial envelope model \citep{SuCook2011}: $\Y_{Si} \mid \X=  \e_{Si}$ and
\begin{eqnarray}
\Y_{Di}\mid(\X_{i}, \Y_{Si})  & = & \alphabf_0 + \Phibf\etabf\X_{i} + \phibf_{D|S}\Y_{Si}  + \e_{D|Si},\label{envalpha2} \\
\Sigmabf_{D|S}  & = & \Phibf \Omegabf \Phibf^{T} + \Phibf_{0}\Omegabf_{0}\Phibf_{0}^{T},\nonumber
\end{eqnarray}
where $\Omegabf \in \real{u \times u}$ and $\Omegabf_{0} \in \real {(k-u)\times (k-u)}$ are positive definite matrices.  
The total real parameters in model (\ref{envalpha2}) is $N_{\Ecm}(u) = k + pu + r(r+1)/2$, which reduces to that given previously for model (\ref{cmodelS})--(\ref{cmodelD}) when $u = k$.  The subscript `ecm' is used to indicate selected key quantities that arise from enveloping $\aspc$ in  constrained model (\ref{gmlm}). A basis $\Phibfhat$ for the maximum likelihood estimator $\widehat{\espc}_{\Sigmabfs_{D|S}}(\aspc)$ of $\espc_{\Sigmabfs_{D|S}}(\aspc)$ is constructed as \citep[Ch. 3]{SuCook2011,CookEnv2018},
\begin{eqnarray}
\Phibfhat 
& = &  \arg \min_{\G}\log |\G^{T}\Sbf_{D|(\X,S)}\G| +\log |\G^{T} \Sbf_{D|S}^{-1}\G|, \label{mlealpha}
\end{eqnarray}
where the minimum is  computed over all semi-orthogonal matrices $\G \in \real{k \times u}$ with $u \leq k$. The fully maximized log likelihood  is 
\begin{equation}\label{loglik2}
\Lhat_{u}=c  - \frac{n}{2}\left\{ \log |\T_{S} |  + \log |\Sbf_{D|S} | +\log |\Phibfhat^{T}\Sbf_{D|(\X,S)}\Phibfhat| + \log |\Phibfhat^{T} \Sbf_{D|S}^{-1}\Phibfhat|\right\}.
\end{equation}
where $c = n\log |\W| -(nr/2)(1+ \log(2\pi))$ with the $\log|\W|$ term corresponding to the Jacobian transformation back to the scale of $\Y$.

 Once $\widehat{\Phibf}$ is obtained we get the following envelope estimators for constrained model (\ref{gmlm}). Specifically, we have 
$\betabfhat_{\Ecm} = \U \alphabfhat_{\Ecm}, \alphabfhat_{\Ecm}= \Pbf_{ \widehat{\Phibfs}} \alphabfhat_{\Cm} = \Phibfhat \etabfhat$,  $\alphabfhat_{0} = \Ybar_{D} -  \alphabfhat_{\Ecm}\Xbar - \phibfhat_{D|S}\Ybar_{S},$ where 
$\etabfhat = \Phibfhat^{T} \alphabfhat_{\Cm}$, and $\phibfhat_{D|S} = \Sbf_{D,S}\Sbf_{S}^{-1} - \alphabfhat_{\Ecm}\Sbf_{\X,S}\Sbf_{S}^{-1}$. We also have 
$\Omegabfhat = \Phibfhat^{T}\Sbf_{D|(\X,S)} \Phibfhat\text{ and }\Omegabfhat_{0} = \Phibfhat_{0}^{T}\Sbf_{D|S} \Phibfhat_{0},$ where
$\Sigmabfhat_{D|S} = \Phibfhat\Omegabfhat\Phibfhat^{T} + \Phibfhat_{0}\Omegabfhat_{0}\Phibfhat_{0}^{T}\text{ and }\Sigmabfhat_{S} = \T_{S}$.
The variances $\Sigmabf_{\W}$ and $\Sigmabf$ can be estimated as  indicated in Section~\ref{sec:gmlm}.

The asymptotic variance for $\alphabfhat_{\Ecm}$ can be deduced from \citet[Prop. 1]{SuCook2011} recognizing that in our application $\Y_{S}$ is random, $\X$ is fixed, and the distribution of $\Y_{S} | \X$ is the same as that of the marginal of $\Y_{S}$:
\[
\avar(\sqrt{n}\vecc(\alphabfhat_{\Ecm})) =  \Sigmabf_{\X}^{-1}\otimes \Phibf \Omegabf \Phibf^T + (\etabf^T\otimes \Phibf_0) \M^{\dagger}(\Sigmabf_{\X}) (\etabf \otimes \Phibf^T_0).
\]
 It can be shown that $\avar(\sqrt{n}\vecc(\alphabfhat_{\Ecm})) \leq \avar(\sqrt{n}\vecc(\alphabfhat_{\Cm}))$, so using an envelope in the constrained model always improves estimation asymptotically.
 
 Because $\espc_{\Sigmabfs_{D|S}}(\aspc) \subseteq \real{k}$,  $\espc_{\Sigmabfs}(\bspc) \subseteq \real{r}$ and $k \leq r$ it is reasonable to expect that $\dim\{\espc_{\Sigmabfs_{D|S}}(\aspc)\} \leq \dim\{\espc_{\Sigmabfs}(\bspc)\}$, as we have estimated in many examples.  However, this relationship between the envelope dimension is not guaranteed in general.  The following proposition gives conditions sufficient to bound $\dim\{\espc_{\Sigmabfs_{D|S}}(\aspc)\}$.  
 
 \begin{prop}\label{prop:dims}
 Assume that $\U = (\Gammabf\G, \Gammabf_{0}\G_{0})$, where the $\Gammabf$'s are as defined for model (\ref{envmlm}), and that $\G \in \real{u \times u_{1}}$ and $\G_{0} \in \real{(r-u)\times(k-u_{1})}$ both have full column rank, so $u_{1} \leq u$.  Then $\dim\{\espc_{\Sigmabfs_{D|S}}(\aspc)\} \leq u_{1} \leq \dim\{\espc_{\Sigmabfs}(\bspc)\}$.
 
 \end{prop}
 \subsection{Enveloping $\betabf$}\label{sec:envbeta}


{Estimation of $\betabf = \U\alphabf$ will be of interest in applications where prediction is important or where $\U$ is selected based on convenience, say,  rather than on criteria that facilitate understanding and inference.  For instance, if $\X$ serves to indicate different treatments then plots of the columns of $\betabf$ versus time give a visual comparisons of the treatment profiles. The choice of $\uspc$ is of course relevant to estimation of $\betabf$, but a basis $\U$ is not uniquely determined. While this flexibility has no effect on the maximum likelihood estimators of $\betabf$ under the constrained model (\ref{gmlm}),  it does affect the envelope estimator of $\betabf$.  This raises the issue of selecting a good basis for  the purpose of estimating $\betabf$ via envelopes.}


Consider re-parameterizing $\U$ as $\U\V^{-1}$ and $\alphabf$ as $\V\alphabf$ for  some positive definite matrix $\V \in \real{k \times k}$, giving $\betabf = \U\alphabf = (\U\V^{-1})(\V\alphabf)$.  We could use either $\espc_{\Sigmabfs_{D|S}}(\aspc)$ or $\espc_{\V\Sigmabfs_{D|S}\V^{T}}(\V\aspc)$ to estimate $\betabf$ as $\betabfhat_{\Ecm} = \U\alphabfhat_{\Ecm}$ or, in terms of re-parameterized coordinates  $\V\alphabf$, as $\betabfhat_{\Ecm, \V} = \U\V^{-1}\widehat{(\V\alphabf})_{\Ecm}$.  In general $\betabfhat_{\Ecm} \neq \betabfhat_{\Ecm,\V}$ and we cannot tell which estimator is necessarily better.  
In this section, we show that the envelope estimator of $\betabf$ is invariant under orthogonal re-parameterization, so we only need to consider diagonal re-parameterization: $\betabf = \U\alphabf = (\U\Lambdabf^{-1})(\Lambdabf\alphabf)$, where $\Lambdabf$ is a diagonal matrix with positive diagonal elements.  In growth curve or longitudinal analyses for instance, the columns of $\U$ may correspond to different powers of time, and then it seems  natural to consider rescaling to bring the columns of $\U$ closer to the same scale.


  The following two propositions provide technical tools for demonstrating that the maximum likelihood envelope estimator of $\betabf = \U\alphabf$ is simply $\betabfhat_{\Ecm} = \U \alphabfhat_{\Ecm}$ when $\U$ is semi-orthogonal, where $\alphabfhat_{\Ecm}$ is the envelope estimator of $\alpha$ under the constrained model  (\ref{gmlm}).

\begin{prop}\label{lemma1}
 (a) Let $\spc \subseteq \real{k}$ be a reducing subspace of the symmetric matrix $\M \in \real{k \times k}$, and let $\V \in \real{p \times k}$ be a semi-orthogonal matrix.  Then  $\V\spc$ is a reducing subspace of $\V \M \V^{T}$.
 (b) Let $\dspc \in \real{p}$ be a reducing subspace of $\V\M\V^{T}$.  Then $\V^{T}\dspc$ is a reducing subspace of $\M$.
 \end{prop}
 
\begin{prop}\label{lemma2}
 Let $\espc_{\M}(\spc) \subseteq \real{k}$ be the smallest reducing subspace of the symmetric matrix $\M \in \real{k \times k}$ that contains $\spc \subseteq \real{k}$, and let $\V \in \real{p \times k}$ be a semi-orthogonal matrix.  Then $\V\espc_{\M}(\spc) $ is the smallest reducing subspace of $\V \M \V^{T}$ that contains $\V\spc$; that is, $\V\espc_{\M}(\spc) =  \espc_{\V\M\V^{T}}(\V\spc)$. 

 \end{prop}
  
 These two propositions show that the results of Section~\ref{sec:envalpha} can be used straightforwardly to get the envelope estimator of $\U\alphabf$ when $\U$ is semi-orthognonal. The standard maximum likelihood estimator of $\U\alphabf$ is just $\U\alphabfhat_{\Cm}$ with asymptotic covariance matrix $\U\Sigmabf_{D|S}\U^{T}$.  In consequence, following the rationale at the beginning of Section~\ref{sec:envalpha}, we seek the maximum likelihood estimator of $\espc_{\U\Sigmabfs_{D|S}\U^{T}}(\U\aspc)$,  which by Proposition~\ref{lemma2} is equal to $\U\espc_{\Sigmabfs_{D|S}}(\aspc)$.
 From Proposition~\ref{lemma2},  the maximum likelihood estimator of $\espc_{\U\Sigmabfs_{D|S}\U^{T}}(\U\aspc)$ is
 $\U\widehat{\espc}_{\Sigmabfs_{D|S}}(\aspc)$, which implies that envelope estimator of $\betabf = \U\alphabf$ is $\betabfhat_{\Ecm}=\U\alphabfhat_{\Ecm}$ with asymptotic variance $\U\avar(\sqrt{n}\alphabfhat_{\Ecm})\U^{T}$.  Propositions~\ref{lemma1} and \ref{lemma2} also  suggest how to proceed when re-prameterizing as $\betabf = \U\alphabf = (\U\Obf^{T})(\Obf\alphabf)$, where $\Obf$ is an orthognal matrix and $\U$ is not necessarily orthogonal.  In that case the envelope estimator of $\Obf\alphabf$ is simply $\Obf\alphabf_{\Ecm}$, and so the envelope estimator of $\betabf$ is invariant under orthogonal re-paramterization  of the kind used here.




Thus, to consider constrained model envelope under a linear transformation of $\U$, it suffices to consider a re-scaling transformation. That is, we consider $\betabf = \U\alphabf = (\U\Lambdabf^{-1})(\Lambdabf\alphabf)$, where $\Lambdabf = \diag(1,\lambda_{2},\ldots,\lambda_{k})$. The first diagonal element of $\Lambdabf$ is 1 to ensure identifiability. We follow the general logic of \citet{CookSu2013scaled} in their development of a scaled version of envelope model (\ref{gmlm}). 

Without loss of generality, we cast our discussion of scaling in the context of conditional model (\ref{cmodelD}).  We suppose that there is a scaling of the response $\Y_{D}$ so that the scaled response $\Lambdabf \Y_{D}$ follows an envelope model in $\Lambdabf \alphabf$ with the envelope $\espc_{\Lambdabfs \Sigmabfs_{D|S}\Lambdabfs}(\Lambdabf \aspc)$ having dimension $v$ and semi-orthogonal basis matrix $\Thetabf \in \real{k \times v}$.  Let $(\Thetabf, \Thetabf_{0})$ denote an orthogonal matrix. Then we can parameterize $\Lambdabf \alphabf = \Thetabf \etabf$ and $\Lambdabf \Sigmabf_{D|S}\Lambdabf = \Thetabf \Omegabf\Thetabf^{T}+\Thetabf_{0}\Omegabf \Thetabf_{0}^{T}$.  This setup can also be  viewed equivalently as a rescaling $\U \mapsto \U\Lambdabf^{-1}$ of $\U$, since  $\Lambdabf \Y_{D} = \Lambdabf (\U^{T}\U)^{-1}\U^{T}\Y = (\Lambdabf^{-1}\U^{T}\U\Lambdabf^{-1})^{-1}\Lambdabf^{-1}\U^{T}\Y$. Since $\Lambdabf \Y_{D}$ is unobserved, we now transform back to the original scale for analysis, leading to the marginal model $\Y_{Si} \mid \X =  \e_{Si}$ and conditional model
\begin{eqnarray}
\Y_{Di}\mid(\X_{i}, \Y_{Si})  & = & \alphabf_0 + \Lambdabf^{-1}\Thetabf \etabf \X_{i} + \phibf_{ D|S}\Y_{Si}  + \e_{ D|Si}, \label{LEmodelD}\\
\Sigmabf_{ D|S} &=&\Lambdabf^{-1}( \Thetabf \Omegabf \Thetabf^{T} + \Thetabf_{0} \Omegabf_{0} \Thetabf_{0}^{T})\Lambdabf^{-1} \nonumber.
\end{eqnarray}
The total real parameters in this scaled envelope model is $N_{\Secm}(v) = 2k-1 + pv + r(r+1)/2$, where the subscript `secm' is used to indicate quantities arising from the scaled envelope version of the conditional model.  For identifiability we typically need $N_{\Secm}(v) \leq N_{\Cm}$ or $p(k-v) \geq k-1$.  
 The goal now is to estimate $\alphabf_{0}$, the coefficient matrix $\betabf = \U  \Lambdabf^{-1}\Thetabf \etabf $ and 
$\Sigmabf_{D|S}$, which requires the estimation of several constituent parameters.  

After maximizing the log likelihood over all parameters except $(\Lambdabf, \Thetabf)$ we have
\begin{equation}\label{mlelambda}
(\Lambdabfhat, \Thetabfhat) =  \arg \min_{\A, \G}\log |\G^{T}\A\Sbf_{D|(\X,S)}\A\G| +\log |\G^{T}\A^{-1} \Sbf_{D|S}^{-1}\A^{-1}\G|,
\end{equation}
where the minimum is  computed over all semi-orthogonal matrices $\G \in \real{k \times v}$  and  diagonal matrices $\A = \diag(1,a_{2},\ldots,a_{k})$. Aside from the inner product matrices $\Sbf_{D|(\X,S)}$ and $\Sbf_{D \mid S}^{-1} $ this is the same as the objective function that \citet{CookSu2013scaled} derived for response scaling prior to using model (\ref{envmlm}), which allowed us to adapt their optimization algorithm to handle (\ref{mlelambda}).

Having determined the maximum likelihood estimators $\Lambdabfhat$ and $\Thetabfhat$,  the remaining parameter estimators are $\betabfhat_{\Secm} =  \U \Lambdabfhat^{-1} \Pbf_{\Thetabfhats}\Lambdabfhat \alphabfhat_{\Cm} $, $\alphabfhat = \Lambdabfhat^{-1}\Pbf_{\Thetabfhats}\Lambdabfhat \alphabfhat_{\Cm}$, $\alphabfhat_{0} = \Ybar_{D} - \betabfhat_{\Secm}\Xbar -  \phibfhat_{ D|S} \Ybar_{S}$, where $\etabfhat = \Thetabfhat^T \Lambdabfhat \alphabfhat_{\Cm}$, $\phibfhat_{D|S} = (\Sbf_{D,S} - \alphabfhat\Sbf_{\X,S} ) \Sbf_S^{-1}$. We also have $\Omegabfhat =  \Thetabfhat^{T}\Lambdabfhat  \Sbf_{ D|(\X,S)} \Lambdabfhat \Thetabfhat $, $\Omegabfhat_{0} = \Thetabfhat_{0}^{T} \Lambdabfhat \Sbf_{D|S}\Lambdabfhat \Thetabfhat_{0} $, where $\Sigmabfhat_{D|S} = \Lambdabfhat^{-1}( \Thetabfhat \Omegabfhat \Thetabfhat^{T} +  \Thetabfhat_{0} \Omegabfhat_{0}\Thetabfhat_{0}^{T}) \Lambdabfhat^{-1}$, $\Sigmabfhat_{S} = \T_{S}$.
The variances $\Sigmabf_{\W}$ and $\Sigmabf$ can be estimated as  indicated in Section~\ref{sec:gmlm}.

 This representation of the scaled envelope estimator $\betabfhat_{\Secm} $ shows the construction process.  First the direct-information response is transformed to $\Lambdabfhat \Y_{D}$. The constrained estimator $\Lambdabfhat \alphabfhat_{\Cm}$ and the envelope estimator $ \Pbf_{\Thetabfhats} \Lambdabfhat \alphabfhat_{\Cm}$ are  then determined in the transformed scale.  Next, the estimator is transformed back to the original scale by multiplying by $\Lambdabfhat^{-1}$ to get  $\Lambdabfhat^{-1} \Pbf_{\Thetabfhats}\Lambdabfhat \alphabfhat_{\Cm}$, which is the estimator of $\alphabf$ in the original scale.  Finally, the estimator in the original scale is multiplied by $\U$ to give the scaled envelope estimator of $\betabf$.  In effect, $\Lambdabfhat$ is a similarity transformation to represent $ \Pbf_{\Thetabfhats}$ in the original coordinate system as  $\Lambdabfhat^{-1} \Pbf_{\Thetabfhats}\Lambdabfhat $.

The fully maximized log likelihood is 
\begin{equation}\label{loglik3}
\Lhat_{u}= c - \frac{n}{2}\left\{\log |\T_{S} |  + \log  |\Sbf_{D|S} |  +
 \log |\Thetabfhat^{T}\Lambdabfhat\Sbf_{D|(\X,S)}\Lambdabfhat\Thetabfhat| + \log |\Thetabfhat^{T} \Lambdabfhat^{-1}\Sbf_{D|S}^{-1}\Lambdabfhat^{-1}\Thetabfhat|\right\},
\end{equation}
where $c = n\log|\W| - (nr/2)(1+ \log(2\pi)) $.  To describe the asymptotic variance of $\betabfhat_{\Secm}$, let $\V_{\Secm} $ denote the upper $p k \times pk$ diagonal block of the asymptotic variance
$\V$ given by Proposition 2 from \citet{CookSu2013scaled} with 
$\Sigmabf$ replaced by $\Sigmabf_{D\mid S}$,  $\Gammabf$ by $\Thetabf$ and $\Gammabf_0$ by $\Thetabf_{0}$ and $\Lambdabf$ with $\Lambdabf^{-1}$. Additionally, $\Omegabf$ and $\Omegabf_0$ in the Cook-Su notation are the same as the corresponding quantities in the decomposition of  $\Sigmabf_{D \mid S}$ for model (\ref{LEmodelD}) . Then $\avar(\sqrt{n}\vecc(\betabfhat_{\Secm})) = (\I_{p} \otimes \U) \V_{\Secm} (\I_{p}\otimes \U^T)$.


%


\subsection{ Testing }\label{sec:testing}

{Using the envelope version (\ref{envalpha2}) of constrained model (\ref{gmlm}), we address in Section \ref{test11}  the adequacy of $\U$ through a test on the rows of $\alphabf$ and in Section \ref{sec:colcontrast} we present 
a test of the column of $\alphabf$ to asses the importance of the predictors.  }

\subsubsection{Evaluating the choice of $\U$ by testing rows of $\alphabf$}\label{test11}

Having selected the dimension $u$ of the envelope, we may also want to test if $\U$ is over specified. This can be achieved by testing if individual rows of $\alphabf$ are equal to $0$.  For instance, if $\Ubf^{T}(t) = (1,t,t^{2},t^{3})$ we might wish to test if the cubic term is necessary by testing if the last row of $\alphabf$ is $0$.  
 
 Consider a test that the last $k_{2} \leq k - u$ rows of the $\alphabf = \Phibf \etabf$ in model (\ref{envalpha2})  all equal $0$.  Following \citet{Su2016} and \citet{Zhu2019}, this hypothesis can be  tested by conformably partitioning in (\ref{envalpha2}) $\Phibf = (\Phibf_{1}^{T}, \Phibf_{2}^{T})^{T}$ with $\Phibf_{j} \in \real{k_{j} \times u}, \;j=1,2$, and then testing if $\Phibf_{2} = 0$, so under the null hypothesis 
$\alphabf = ((\Phibf_{1}\etabf)^{T}, 0)^{T}$.  The restriction $k_{2} \leq  k-u$ on the number of rows tested arises because the rank of $\Phibf_{1}$ must equal $u$ under both the null and alternative hypothesis.  If $k_{2} = k - u$ then without loss of generality we can take $\Phibf_{1} = \I_{u}$.

When $k_{2} < k-u$, the maximum likelihood estimator of $\Phibf_1$ can be found by following the steps leading to  (\ref{mlealpha}) and then introducing the restriction that  $\G = (\G_{1}^{T}, 0)^{T}$.  Partition $\Y_{D} = (\Y_{D_{1}}^{T}, \Y_{D_{2}}^{T})^{T}$ to conform to the partitioning of $\Phibf$.  Then
\begin{equation}\label{likpart40}
\Phibfhat _1=  \arg \min_{\G_{1}} \log |\G_{1}^{T}\Sbf_{D_{1}|(\X,S)} \G_{1}| +\log |\G_{1}^{T} \Sbf_{D_{1}|(D_{2},S)}^{-1} \G_{1}|,
\end{equation}
where the minimum is computed over all semi-orthogonal matrices $\G_{1}\in \real{k_{1} \times u}$, $\Sbf_{D_{1}|(D_{2},S)} $ is the sample residual covariance matrix of the regression  of 
$\Y_{D_{1}} $ on $(\Y_{D_{2}}, \Y_{S})$ with an intercept, and $\Sbf_{D_{1}|(\X,S)}$ is the sample residual covariance matrix of the regression  of 
$\Y_{D_{1}} $ on $ (\X,\Y_{S})$ with an intercept.  When $k_{2} = k-u$, we must have $\Phibf_{1} = \I_{u}$ and no estimate of $\Phibf_{1}$ is required.

The likelihood ratio test statistic is computed as twice the difference between the log likelihood under the null hypothesis,
\begin{equation}\label{loglik4}
\Lhat_{u,k_{2}}= c - \frac{n}{2}\left\{\log |\T_{S} |  + \log |\Sbf_{D_{2}|S}|  + \log |\Sbf_{D_{1}|(D_{2}, S})| +\log |\Phibfhat_{1}^{T}\Sbf_{D_{1}|(\X,S)}\Phibfhat_{1}| + \log |\Phibfhat_{1}^{T} \Sbf_{D_{1}|(D_{2},S)}^{-1}\Phibfhat_{1}|\right\},
\end{equation}
 and the fully maximized log likelihood (\ref{loglik2}). Here $c = n \log |\W|-(nr/2)(1+ \log(2\pi))$. Under the null hypothesis $\Phibf_{2} = 0$ this difference is asymptotically distributed as a chi-squared random variable with $uk_{2}$ degrees of freedom.

\subsubsection{Evaluating predictors by testing column contrast of $\alphabf$}\label{sec:colcontrast}

In some studies we may wish to estimate and infer about column contrasts $\alphabf_{1} = \alphabf \cbf_{1}$, where $\cbf_{1} \in \real{p \times p_{1}}$ is a user-selected matrix of known constants with $p_{1} < p$. For instance, when $\X$ is a treatment indicator, testing column contrasts allows testing equality of treatment means. Let $\aspc_{1} = \spn(\alphabf_{1})$.  We could use the conditional model (\ref{gmlm}), basing estimation and inference on $ \alphabfhat_{\Cm} \cbf_{1}$.  Or we could proceed following the envelope analysis of Section~\ref{sec:envalpha} and use the estimator $\alphabfhat_{\Ecm} \cbf_{1}$ with asymptotic variance $\avar(\sqrt{n}\vecc(\alphabfhat_{\Ecm}\cbf_{1}))$
as a basis for inference.  {The latter estimator is preferable since $\avar(\sqrt{n}\vecc(\alphabfhat_{\Ecm}\cbf_{1})) \leq \avar(\sqrt{n}\vecc(\alphabfhat_{\Cm}\cbf_{1}))$.  But there is a potential to gain  additional asymptotic efficiency by using envelope methods to estimate $\alphabf_{1}$ directly.}

To develope an  envelope estimator of $\alphabf_{1}$, we first parameterize model (\ref{gmlm}) so $\alphabf_{1}$ appears explicitly.  Select a matrix $\cbf_{2} \in \real{p \times p_{2}}$, $p_{1} + p_{2} = p$, so that $\C = (\cbf_{1}, \cbf_{2}) \in \real{ p \times p}$ is non-singular and define new predictors and parameters as $\Z = \C^{-1}\X$ and $\alphabf_{2} = \alphabf\cbf_{2}$. Then we have 
\[
\U\alphabf \X = \U\alphabf\C\C^{-1}\X =\U(\alphabf_{1},\alphabf_{2})\Z = \U\alphabf_{1}\Z_{1} + \U\alphabf_{2}\Z_{2},
\]
where the row partitioning of $\Z = (\Z_{1}^{T}, \Z_{2}^{T})^{T}$ conforms to the column partitioning of $\C$.  Following the logic used previously in this section, we obtain the marginal and conditional models: $\Y_{Si} \mid \Z =  \e_{Si}$ and
\begin{eqnarray}
\Y_{Di}\mid(\X_{i}, \Y_{Si})  & = & \alphabf_0 + \alphabf_{1} \Z_{1i} + \alphabf_{2} \Z_{2i} + \phibf_{D|S}\Y_{Si}  + \e_{D|Si},\nonumber  \\
& = &  \alphabf_0 + \alphabf_{1} \Z_{1i} + \omegabf \K_{i} + \e_{D|Si}, \label{pcmodelD}
\end{eqnarray}
where $\omegabf = (\alphabf_{2}, \phibf_{D|S})$, $\K_{i} = (\Z_{2i}^{T},\Y_{Si}^{T})^{T}$ and the other terms  are as defined previously.  This model is of the same form as (\ref{cmodelD}) and so a semi-orthogonal basis $\Phibf \in \real{k \times u_{1}}$ for $\espc_{\Sigmabfs_{D|S}}(\aspc_{1})$ with dimension $u_{1} \leq k$ can be incorporated into model (\ref{pcmodelD}) as follows. $\Y_{Si} \mid \Z =  \e_{Si}$ and
\begin{eqnarray}
\Y_{Di}\mid(\X_{i}, \Y_{Si})  
& = &  \alphabf_0 + \Phibf\etabf \Z_{1i} + \omegabf \K_{i} + \e_{D|Si}, \label{pcmodelD1}\\
\Sigmabf_{D|S} & = & \Phibf \Omegabf \Phibf + \Phibf_{0} \Omegabf_{0} \Phibf_{0}, \nonumber
\end{eqnarray}
where $\Sigmabf_{S}$ and $\Sigmabf_{D|S}$ are as defined in Section~\ref{sec:gmlm}.  The number of free real parameters in this model is $k + u_{1}p_{1} + p_{2}k + r(r+1)/2$.
The envelope estimators can now be obtained straightforwardly by following the steps in Section~\ref{sec:envalpha}, and the asymptotic variance of envelope estimator $\alphabfhat_{1} = \Pbf_{\Phibfhats}\alphabfhat_{\Cm}\cbf_{1}$ of $\alphabf_{1}$ is 
\[
\avar(\sqrt{n}\vecc(\alphabfhat_{1})) =   \Sigmabf_{\Z_{1}|\Z_{2}}^{-1}\otimes \Phibf \Omegabf \Phibf^T + (\etabf^T\otimes \Phibf_0) \M^{\dagger}( \Sigmabf_{\Z_{1}|\Z_{2}} ) (\etabf \otimes \Phibf^T_0),
\]
where $\Sigmabf_{\Z_{1}|\Z_{2}} = \lim_{n\rightarrow \infty} \Sbf_{\Z_{1}|\Z_{2}}$.

Looking ahead and adapting the discussion of Section~\ref{sec:Balpha}, the envelope estimator of $\U\alphabf_{1}$ is simply $\U\alphabfhat_{1}$ with asymptotic variance 
\begin{equation}\label{varUalpha1}
\avar(\sqrt{n}\;\vecc(\U\alphabfhat_{1})) = (\I_{p_{1}}\otimes \U)\avar(\sqrt{n}\;\vecc(\alphabfhat_{1}))(\I_{p_{1}}\otimes \U^{T}),
\end{equation}
where $\avar(\sqrt{n}\vecc(\alphabfhat_{1}))$ is as given previously.

These results can be adapted to obtain an envelope estimator of the average profile $E(\Y|\X_{\new}) = \U\alphabf_{0} + \U\alphabf\X_{\new}$ at a new value $\X_{\new}$ of $\X$ by setting $\cbf_{1} = \X_{\new}$, so $\alphabf_{1} = \alphabf \X_{\new}$.  Assuming without loss of generality that $\X$ and $\Y_{S}$ in (\ref{pcmodelD1}) are centered, it follows that $\alphabfhat_{0} = \Ybar_{D}$ and thus $\widehat{\U\alphabf_{0}} = \Pbf_{\U}\Ybar$.  The estimator of the average profile is then
$
\widehat{E}(\Y|\X_{\new}) = \Pbf_{\U}\Ybar + \U \alphabfhat_{1} =  \Pbf_{\U}\Ybar + \U \Pbf_{\Phibfhats}\alphabfhat_{\Cm}\X_{\new}.
$
Since $\Ybar$ and $\alphabfhat_{1} $ are asymptotically independent, we get the asymptotic variance
$
\avar(\sqrt{n}\;\widehat{E}(\Y|\X_{\new}) ) = \Pbf_{\U}\Sigmabf \Pbf_{\U} +\avar(\sqrt{n} \;\vecc(\U\alphabfhat_{1})),
$
where the second addend on the right hand side is given by (\ref{varUalpha1}).  This envelope estimator has the potential to be substantially less variable than plugin estimators mentioned at the beginning of Section~\ref{sec:colcontrast}.  A potential disadvantage is that a new envelope estimator is required for each profile determined by the value of $\X_{\new}$.

{ \subsection{Estimation under model (\ref{igmlm})}\label{sec:estalpha1.3}

The modifications necessary to adapt the results in Sections~\ref{sec:envalpha}--\ref{sec:testing} for  model (\ref{igmlm}) all stem from the new model for the subordinate response, $\Y_{S} = \W_{2}^{T}\betabf_{0} + \e_{S}$, and the new definitions of $ \alphabf_{0} = \W_{1}^{T}\betabf_{0}-\phibf_{D|S}\W_{2}^{T}\betabf_{0}$ for models (\ref{envalpha2}),  (\ref{LEmodelD}) and (\ref{pcmodelD1}).  This implies that $\T_{S}$ is replaced by $\Sbf_{S}$ throughout, including log likelihoods (\ref{loglik2}),  (\ref{loglik3}) and (\ref{loglik4}), and that the estimator of $\betabf_{0}$ can be constructed as indicated near the end of Section~\ref{sec:gmlm}.  There is no change in the objective functions (\ref{mlealpha}), (\ref{mlelambda}) and (\ref{likpart40}), and consequently no change in the envelope estimators of $\alphabf$ and $\betabf$ or their asymptotic variances.}

\section{Simulations}\label{simulations}
 \subsection{Efficiency Comparison between envelope and constrained estimator}\label{subsec: sim_eff}
 We first evaluate the efficiency of  the envelope estimator $\widehat\betabf_{\Em}$ and the constrained estimator $\widehat\betabf_{\Cm}$ using simulations in two scenarios. We also include the unconstrained estimator $\widehat\betabf_{\Um}$ as a reference. In  
 Scenario 1, the eigenvalue corresponds to the material part is small relative to the immaterial part and the dimension of $\U$ is large; thus the envelope estimator is expected to have substantial efficiency gain. In Scenario 2, the eigenvalue of the immaterial part is small relative to that of the material part and the envelope estimator is not expected to have substantial efficiency gain. The simulation for Scenario 1 is carried out in the following steps.

 \begin{enumerate}[Step 1.]
{\color{black}{		\item We first generated a sample of size $n=5000$. For each individual $i$, we generated $p=8$ predictors $\X_i$ from a multivariate normal distribution with mean 0 and variance $\C\C^T$, where each element in $\C$ is  identically and independently distributed with a standard normal distribution $N(0,1)$.  {\color{blue}  Comment: The editor said that ``The simulations should included cases where predictors are subject to substantial  dependency.'' I'm not sure this qualifies as substantial dependency since the expected covariance is 0.  Perhaps mentioning the distribution of the predictor correlations in a typical simulation would do.  Alternatively, change the generation scheme to be compound symmetric so the correlations can be specified easily, say .8.  Or perhaps one with a small correlation say .5 and one with a larger correlation say .85}
	\item Set $r=20$, $u=6$, $q=15$, $q_1=4$ and $q_2=q-q_1$. Set $\Omegabf=\bdiag(0.5\I_{u-q_1},1.5\I_{q_1})$ and $\Omegabf_0=50\I_{r-u}$. Set $(\Gammabf,\Gammabf_0)=\Obf$ and let $\Sigmabf = \Gammabf \Omegabf \Gammabf^T + \Gammabf_0  \Omegabf_0 \Gammabf_0^T$, where $\Obf$ is an orthogonal matrix obtained by singular value decomposition of a randomly generated matrix. Set $\etabf=\K_1\K_2$, where $\K_1\in\mathbb{R}^{u\times q_1}$, $\K_2\in\mathbb{R}^{ q_1\times p}$, each element in $\K_1$ and $\K_2$ is identically and independently generated from $N(0,1)$. Set $\betabf=\Gammabf\etabf$.  Let $\U=(\Gammabf,\Gammabf_0)\Phibf$, where $\Phibf=\bdiag\{\M^{\U},\M_0^{\U}\}$, $\M^{\U}=\K_1$ and $\M_0^{\U}=(\I_{q_2},\mathbf{0}_{q_2\times (r-u-q_2)})^T$.

}}

	\item For each individual $i$, generate $\Y_i$ identically and independently from normal distribution $N(\betabf\X_i,\Sigmabf)$.
	\item Calculate $\widehat\betabf_{\Um}$, $\widehat\betabf_{\Em}$ and $\widehat\betabf_{\Cm}$, where $\U$ is correctly specified when calculating $\widehat\betabf_{\Cm}$.
	\item Repeat Steps 3--4 100 times.
\end{enumerate}

\subsubsection{Scenario 1}
\begin{figure}[!h]
	
	\caption{Box plot of $\widehat\betabf_{\Um}-\betabf$, $\widehat\betabf_{\Em}-\betabf$ and $\widehat\betabf_{\Cm}-\betabf$ in two scenarios in 100 simulations.}
	\centering
	\begin{subfigure}[b]{0.45\textwidth}
		\includegraphics[width=\textwidth]{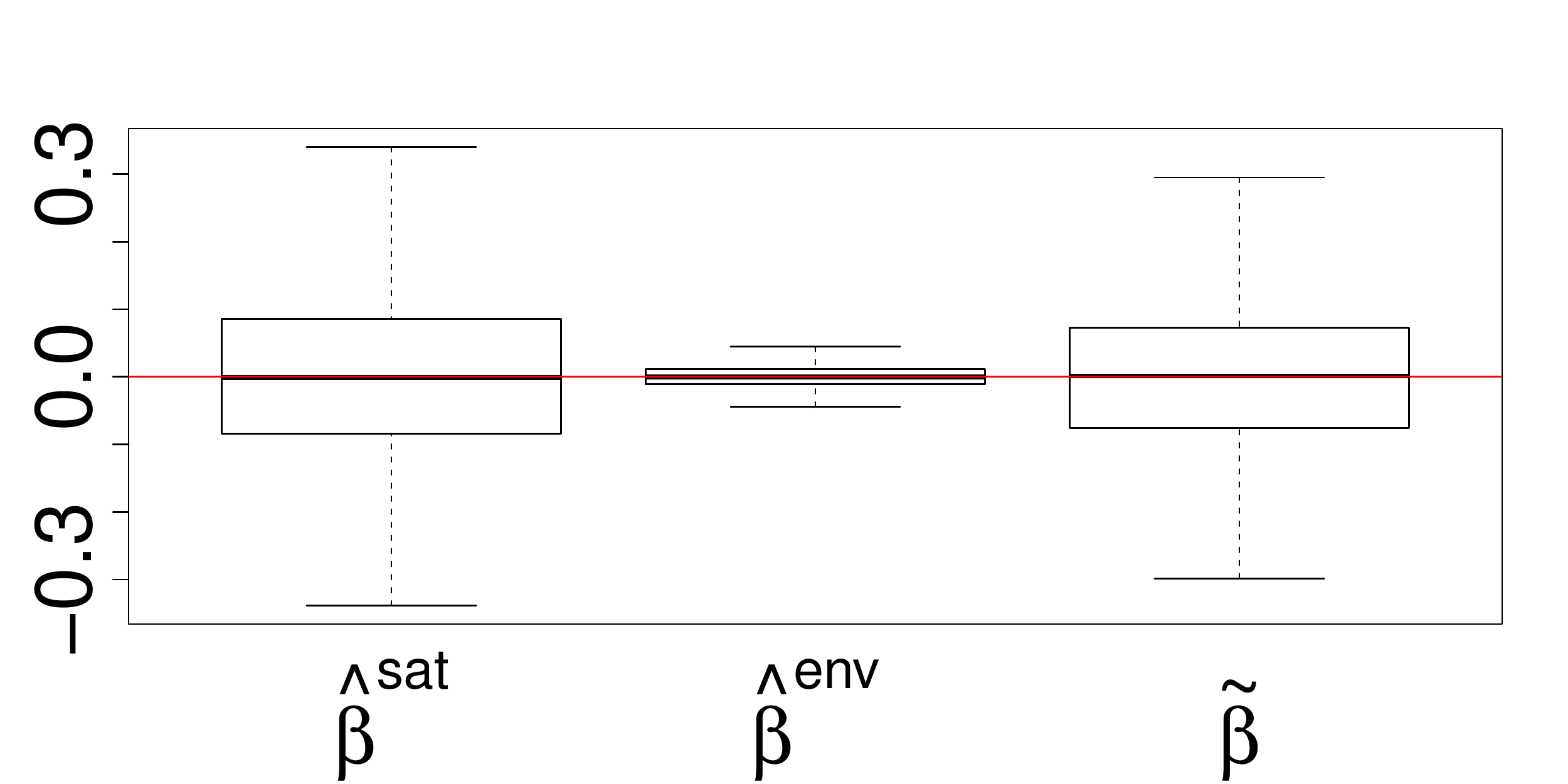}
		\caption{Scenario 1}
		\label{fig: sim_eff_env_win}
	\end{subfigure}\hspace{10mm}
	\begin{subfigure}[b]{0.45\textwidth}
		\includegraphics[width=\textwidth]{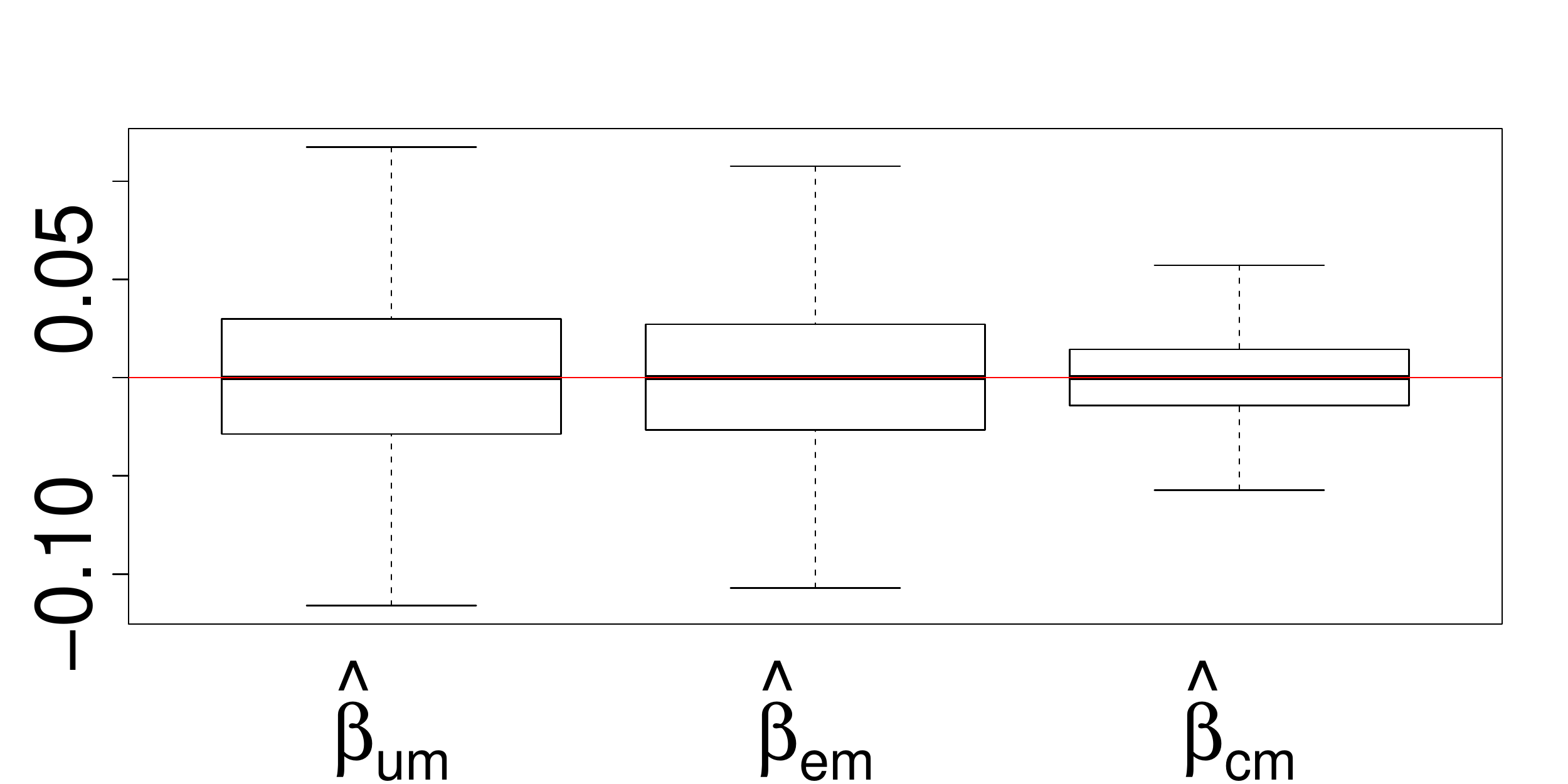}
		\caption{Scenario 2}
		\label{fig: sim_eff_tilde_beta_win}
	\end{subfigure}
	\label{fig: sim_eff}
	\end{figure}

From the choice of $\etabf$ in Step 2, we have $\colrank(\etabf)=q_1$, and $\spant(\betabf)$ is strictly contained in both $\spant(\Gammabf)$ and $\spant(\U)$ since the dimension of $\spant(\betabf)$ is $q_1=4$ which is smaller than $\min(u,q)=6$. Specifically, we also have $\spant(\betabf)=\spant(\Gammabf^{\U})=\spant(\Gammabf)\cap \spant(\U)$. Its easy to see that $\alphabf=(\K_2^T,\zerobf_{p\times q_2})^T$ in this example. Among the 100 simulations, the envelope dimension was always correctly estimated as 6 using BIC. The empirical result of $\widehat\betabf_{\Um}-\betabf$, $\widehat\betabf_{\Em}-\betabf$ and $\widehat\betabf_{\Cm}-\betabf$ are shown in Figure \ref{fig: sim_eff_env_win}, where all the elements of $\betabf$ are plotted in the same boxplot as if they are from the same population and the outliers are suppressed for a cleaner representation. Since $\U$ is correctly specified, $\widehat\betabf_{\Cm}$ is an asymptotically unbiased estimator as are $\widehat\betabf_{\Um}$ and $\widehat\betabf_{\Em}$.  Hence, the boxplot of three estimators are all centered at 0. In Step 2, the larger eigenvalues of $\Sigmabf$ are contained in $\Omegabf_0$ rather than $\Omegabf$. That is, the variability of the immaterial part is bigger than that of the material part. Additionally, the column space of $\U$ is very conservatively specified as $q=15$, which is much bigger than the dimension of $q_1=\colrank(\betabf)=4$  and the $\spant(\U)$ contains 11 eigenvectors corresponds to large eigenvalues (i.e., 50 in this simulation). Hence, this scenario is in favor of the envelope estimator in terms of the efficiency. Indeed, the envelope estimator is the most efficient estimator among the three estimators, while $\widehat\betabf_{\Cm}$ is also more efficient than the {\color{blue}unconstrained} estimator $\widehat\betabf_{\Um}$. 

{\color{black}{The average estimated asymptotic variances were close to the theoretical asymptotic variances calculated using the true parameter values for all three estimators. The mean of the theoretical asymptotic variances across all the elements in three estimators are 127.22 for $\sqrt{n}\widehat\betabf_{\Um}$ and 99.75 for $\sqrt{n}\widehat\betabf_{\Cm}$ but is only 1.70 for $\sqrt{n}\widehat\betabf_{\Em}$.  That is, in this setting, the envelope estimator is about 58 times more efficient that the {\color{blue} constrained estimator and 75   times more efficient than  the unconstrained estimator.} 
}}

\subsubsection{Scenario 2}
To carry out simulations in Scenario 2, we modify Step 2 above as $q=6$, $\Omegabf=\bdiag(50\I_{u-q_1},0.5\I_{q_1})$ and $\Omegabf_0=0.5\I_{r-u}$. In this scenario, the larger eigenvalues of $\Sigmabf$ are associated with  $\Omegabf$, and the dimension of $\U$ can be seen to be  just 2 dimensional larger than the true dimension of $\betabf$.  Hence, the envelope method is at a disadvantage in terms of the efficiency as compared with $\widehat\betabf_{\Cm}$. In the 100 simulations, the envelope dimension is again always correctly estimated as 6. The empirical biases of the envelope and $\widehat\betabf_{\Cm}$ are shown in Figure \ref{fig: sim_eff_tilde_beta_win}. Again, all three estimators are centered around 0, indicating the asymptotic unbiasedness. As expected, the estimator $\widehat\betabf_{\Cm}$ is the most efficient among the three estimators, while the envelope estimator $\widehat\betabf_{\Em}$ is still more efficient than the {\color{blue} unconstrained} estimator $\widehat\betabf_{\Um}$. 


{\color{black}{The average estimated asymptotic variance of the three estimators were all close to their theoretical values. 
The average empirical variances of all the elements in three estimators  are 20.68 for $\sqrt{n}\widehat\betabf_{\Um}$, 19.53 for $\sqrt{n}\widehat\betabf_{\Em}$ and 4.87 for $\sqrt{n}\widehat\betabf_{\Cm}$.  That is, in this setting, the estimator using a correctly specified $\U$ is on average about 4 times of more efficient that the {\color{blue} unconstrained} estimator and the envelope estimator.}}

{\color{black}{\subsection{Potential Bias of the constrained estimator}\label{sec: sim_bias}

We conducted a small simulation, generating data from envelope model (\ref{envmlm}),  to further  illustrate potential bias effects. The sample size and parameters are chosen the same as in Section \ref{subsec: sim_eff}.  The sample size was taken to be large so the bias effects might be clear. It is known that the efficiency gains from fitting (\ref{envmlm}) will be much greater in Scenario 1 than in Scenario 2. 
\begin{figure}[ht!]
    \centering
        \includegraphics[scale=.4]{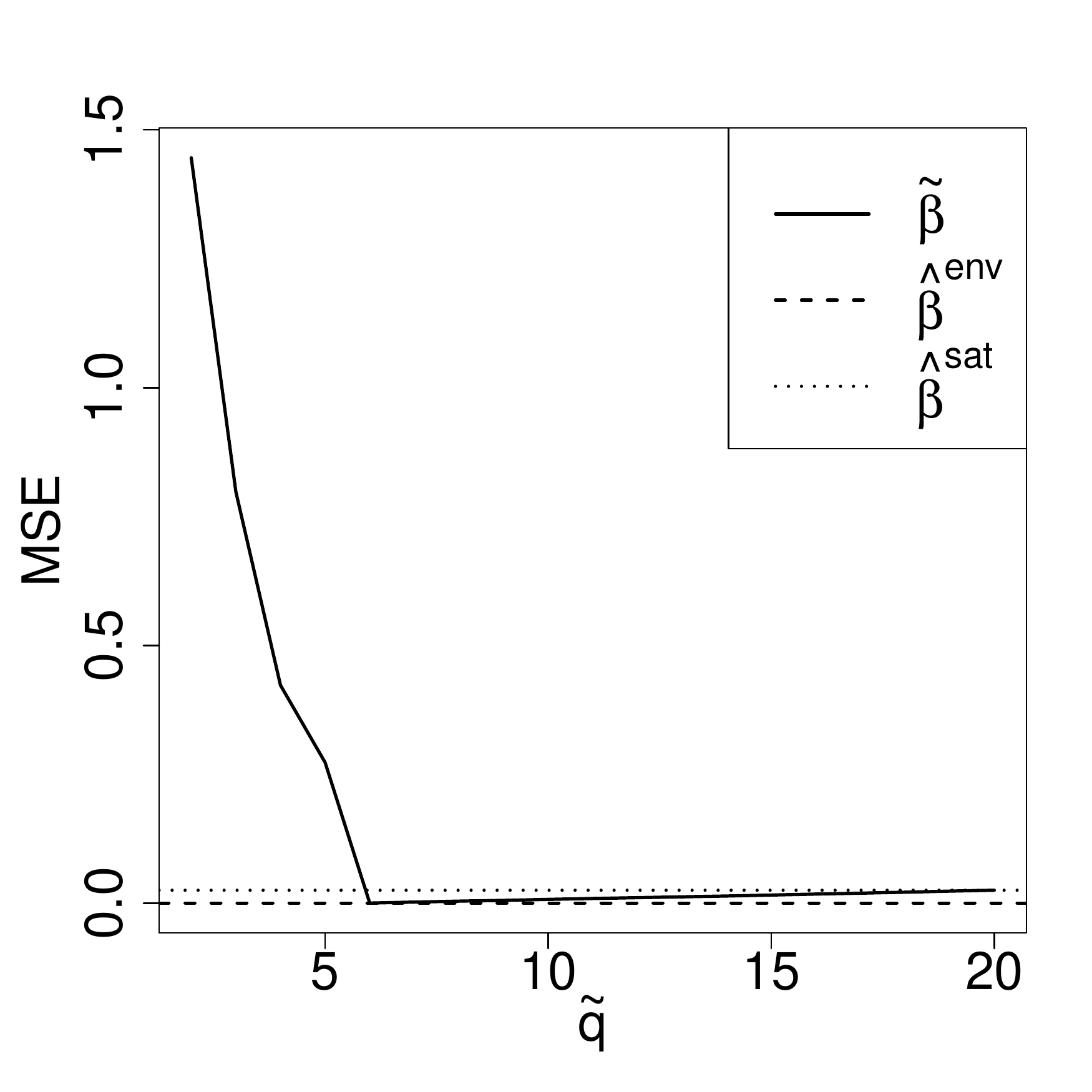}
        \caption{Illustration of potential bias in the constrained estimator (\ref{igmlm}) under Scenario 1, where $k = \dim(\uspc)$, $\uspc = \spn\{(\Gammabf, \Gammabf_{0}) (\I_{k}, 0)^{T}\}$ and MSE denotes the average element-wise squared error  for the indicated estimators. The lines for $\betabfhat_{\Em}$ and $\betabfhat_{\Um}$ are indistinguishable.}
        \label{fig:bias}
\end{figure}

Response vectors were then generated according to model (\ref{envmlm}) using normal errors and the resulting data fitted to obtain the envelope estimator $\betabfhat_{\Em}$.  We used the same data to construct the unconstrained estimator $\betabfhat_{\Um}$ and the constrained estimator $\betabfhat_{\Cm}$ with different selections for $\U = (\Gammabf, \Gammabf_{0}) \A_{k}$ where $\A_{k} = (\I_{k}, 0)^{T}$, $k=1,\ldots,r$.  For $k < u$, $\bspc \not \subseteq \uspc$ and so $\betabfhat_{\Cm}$ is biased.  But for $k\geq u$, $\bspc \subseteq \uspc$ and there is no bias in $\betabfhat_{\Cm}$.  We summarized the bias by computing the mean squared error over all elements $\beta_{ij}$ of $\betabf$: $\mathrm{MSE} = (rp)^{-1} \sum_{i=1}^{r}\sum_{j=1}^{p}(\betahat_{(\cdot),ij} - \beta_{ij})^{2}$ for the three estimators $\betabfhat_{\Um}$, $\betabfhat_{\Cm}$ and  $\betabfhat_{\Em}$.  Shown in Figure~\ref{fig:bias} are plots of the MSE averaged over 100 replications of this scheme for Scenario 1,  each replication starting with the generation of the response vectors.  The constant MSE for  $\betabfhat_{\Em}$ was $3e^{-4}$ and that for unconstrained model was about $8$ times greater at $2.5e^{-3}$. The MSE for the constrained estimator decreased monotonically from its maximum value  $1.74$ at $k=1$ to its minimum value, which was around $3e^{-4}$, at $k=u=6$ and then increased monotonically to  $0.03$ at $k=20$.  The corresponding plot for Scenario 2 is graphically indistinguishable and so is not presented.  It seems clear that the bias in the constrained estimator can be substantial until we achieve $\bspc \subseteq \uspc$, at which point the three estimators become indistinguishable on the scale of Figure~\ref{fig:bias}.

}}

{\color{black}{

{Assuming that $\U$ is correctly specified, we imposed the envelope structure on $\alphabf$ and referred to the new envelope estimator as $\widehat\betabf_{\Ecm}$. We carred out the simulations similar to those in  Section \ref{subsec: sim_eff}, replacing Steps 2--4 with the following.}

 \begin{enumerate}[Step 2*.]
	\item[Step 2*.] Set $r=20$, $u^{*}=3$, $q=15$. Set $\Omegabf^{*}=0.5\I_{u^{*}}$ and $\Omegabf_0=50\I_{q-u^{*}}$. Set $(\Gammabf^{*},\Gammabf^{*}_0)=\I$ and let $\var(\varepsilonbf_{\D|\Sbf})=\Sigmabf_{\D|\Sbf} = \Gammabf^{*} \Omegabf ^{*}{\Gammabf^{*}}^T + \Gammabf_0^{*}  \Omegabf_0^{*} {\Gammabf_0^{*}}^T$. Generate $\etabf^{*}\in\real{u^{*}\times p}$ and $\U$, where each element in $\etabf^{*}$ and $\U$ is identically and independently generated from $N(0,1)$.  Set $\alphabf^{*}=\Gammabf^{*}\etabf^{*}$ and $\betabf^{*}=\U\alphabf^{*}$. 

	\item[Step 3*.] For each individual $i$, generate $\Y_{\Sbf i}$ identically and independently from normal distribution $N(\zerobf,\I_{r-q})$. Generate $\Phibf\in\real{q\times (r-q)}$, where each element is generated identically and independently from standard normal. Generate $\Y_{\D i}$ from distribution $N(\alphabf^{*}\Z_i+\Phibf\Y_{\Sbf i},\Sigmabf_{\D|\Sbf})$
	\item[Step 4*.] Calculate  $\widehat\betabf_{\Cm}$ and  $\widehat\betabf_{\Ecm}$, where $\U$ is correctly specified for both estimators. 
\end{enumerate}

The average MSE of $\widehat\betabf_{\Cm}$ and $\widehat\betabf_{\Ecm}$ was $4e^{-3}$ and $1e^{-3}$. The Monte Carlo mean variances over all the elements were 21.76 and 5.27 for $\sqrt{n}\tilde
\betabf$ and $\sqrt{n}\widehat\betabf_{\Ecm}$, demonstrating the efficiency of the additional envelope structure over the $\widehat\betabf_{\Cm}$ estimator.
}}

\section{Applications}\label{sec:applications}

\subsection{Dental data revisited}\label{sec:dental}

The dental data consists of measurements of the distance (mm) from the center of the pituitary to the pterygomaxillary fissure for each of 11 girls and 16 boys at ages 8, 10, 12, and 14 years ($t$). Since their introduction by \cite{PotthoffRoy1964}, these data have been  used frequently to illustrate the analysis  of longitudinal data.  We respect  that tradition in this section.  We removed the outlying and influential male case described by   \cite{PanFang2002}  prior to application of the methods discussed herein.  We set the goal to characterize the differences between boys and girls rather than profile modeling and so we contrasted the behavior of estimators from the unconstrained model (\ref{mlm}), the envelope  model (\ref{envmlm}), the constrained model (\ref{igmlm}) and the  envelope version of model (\ref{igmlm}) discussed in Section~\ref{sec:envalpha}.

Consistent with the literature, we fitted constrained model (\ref{igmlm}) and its envelope counterpart with the rows of $\U$ being $\Ubf^{T}(t) = (1,t)$.  The estimated dimension of the envelope for model (\ref{envmlm}) was $u=2$, and thus it was inferred that only two linear combinations of the response vectors are needed to fully characterize the differences between boys and girls. The estimated dimension of the envelope for the constrained envelope model (\ref{envalpha2}) was $u=1$.
 Table~\ref{tab:dental} shows the estimated asymptotic variances, determined by the plug-in method, for the four estimators  $\betabfhat_{\Um}$, $\betabfhat_{\Em}$,  $\betabfhat_{\Cm}$ and $\betabfhat_{\Ecm}$.  The unconstrained model has the worst estimated performance, followed by the regular envelope model and the constrained model.  The enveloping in the constrained model has the best estimated performance. We would need to increase the sample size by about 2.5 times for the constrained estimator $\betabfhat_{\Cm}$ to have the performance estimated for the enveloped version $\betabfhat_{\Ecm}$ with the current sample size.

\begin{table}[ht]
\centering
\caption{Estimated asymptotic variances $\avar(\sqrt{n}\betabfhat_{(\cdot)})$ of the four elements of  $\betabfhat_{\Um}$ from the unconstrained model (\ref{mlm}), $\betabfhat_{\Em}$ from the envelope model  (\ref{envmlm}), $\betabfhat_{\Cm}$ from the constrained model (\ref{igmlm}) and $\betabfhat_{\Ecm}$ from the envelope version of constrained model (\ref{igmlm}).}\label{tab:dental}

\bigskip
\begin{tabular}{lrrrr}
& & Age & & \\
  \hline
 $\betabfhat_{(\cdot)}$ & $8\;\;\;$ & $10\;\;$  & $12\;\;$  & $14\;\;$  \\ 
  \hline
$\betabfhat_{\Um}$ & 15.53 & 16.41 & 25.42 & 18.95 \\ 
 $\betabfhat_{\Em}$ & 15.29 & 13.56 & 22.79 & 18.73 \\ 
$\betabfhat_{\Cm}$ & 13.97 & 13.57 & 15.00 & 18.27 \\ 
 $\betabfhat_{\Ecm}$  & 5.88 & 9.16 & 13.16 & 17.89 \\ 
   \hline
\end{tabular}
\end{table}

The relatively bland performance of the envelope estimator $\betabfhat_{\Em}$ can be traced back to the estimated eigen-structure of $\Sigmabf$.  The eigenvalues of $\Omegabfhat$ and $\Omegabfhat_{0}$ were $(14.61, 1.10)$ and $(2.20, 0.70)$.  Envelopes offer relatively little gain when most of the variation in the response is associated material information, as is the case here.  On the other hand, the eigenvalues of $\Omegabfhat$ and $\Omegabfhat_{0}$ arising from enveloping in the constrained model were $0.02$ and $8.31$.  In this case most of the variation in the direct response $\Y_{D}$  is associated with immaterial information, the general setting when envelopes perform well.

\begin{figure}[ht!]
\centerline{\hfill
\includegraphics[width=2.8in]{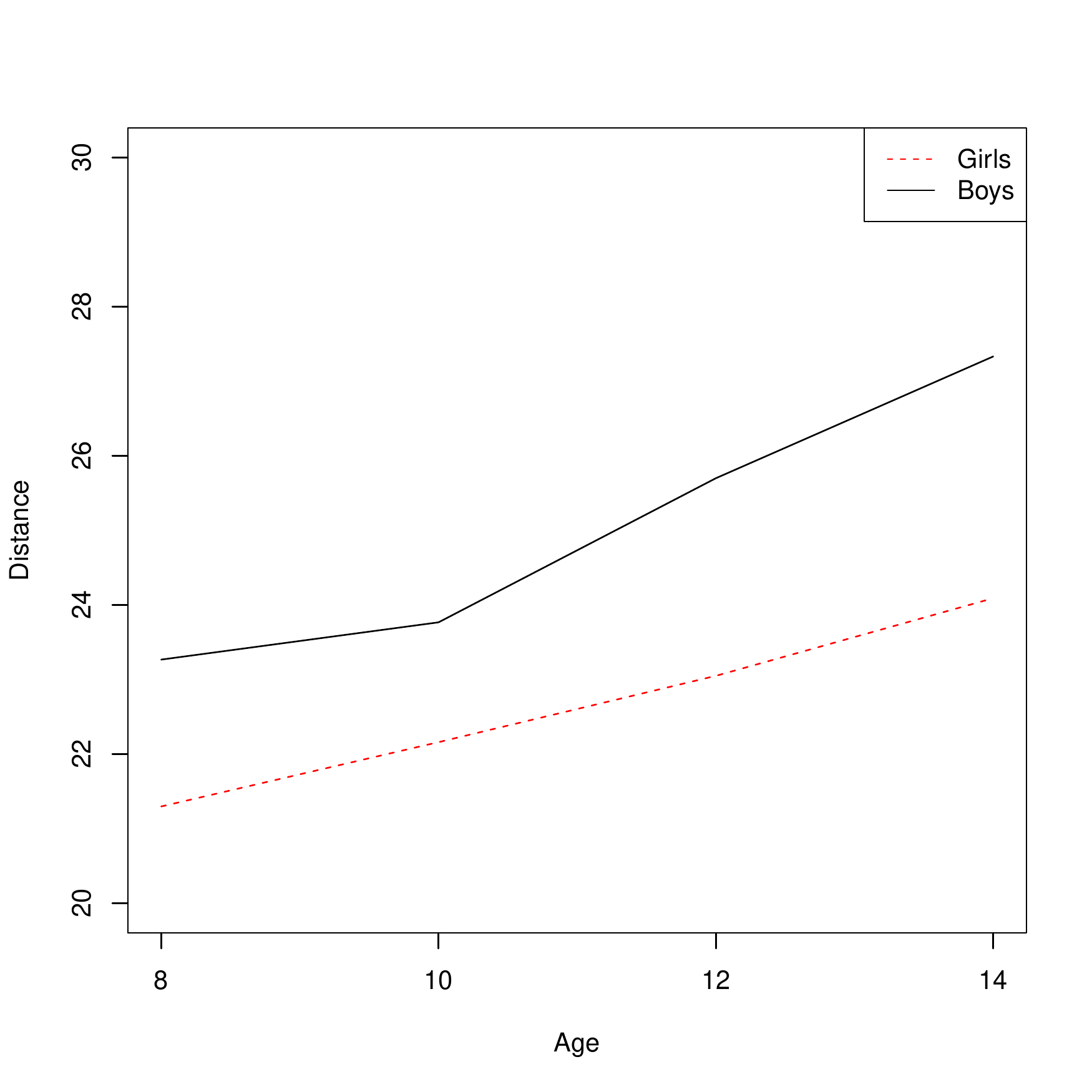}
\hfill
\includegraphics[width=2.8in]{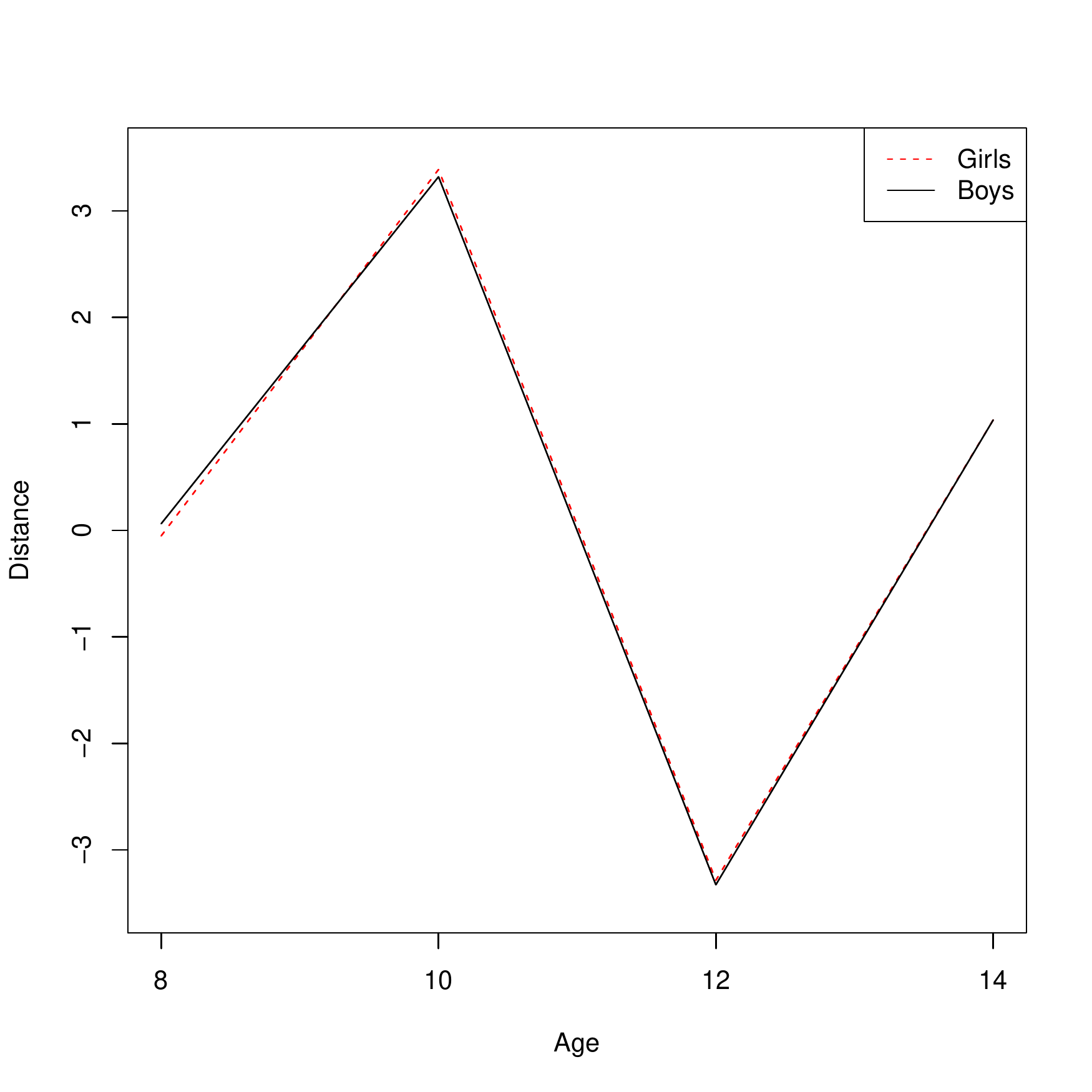}
\hfill}
\centerline{\hfill\makebox[2.1in]{a. Fitted values $\Yhat_{\Em}$ from (\ref{envmlm})}
\hfill\makebox[2.1in]{b. Projected means $\Q_{\Gammabfhats} \Ybar$}\hfill}
\centerline{\hfill
\includegraphics[width=2.8in]{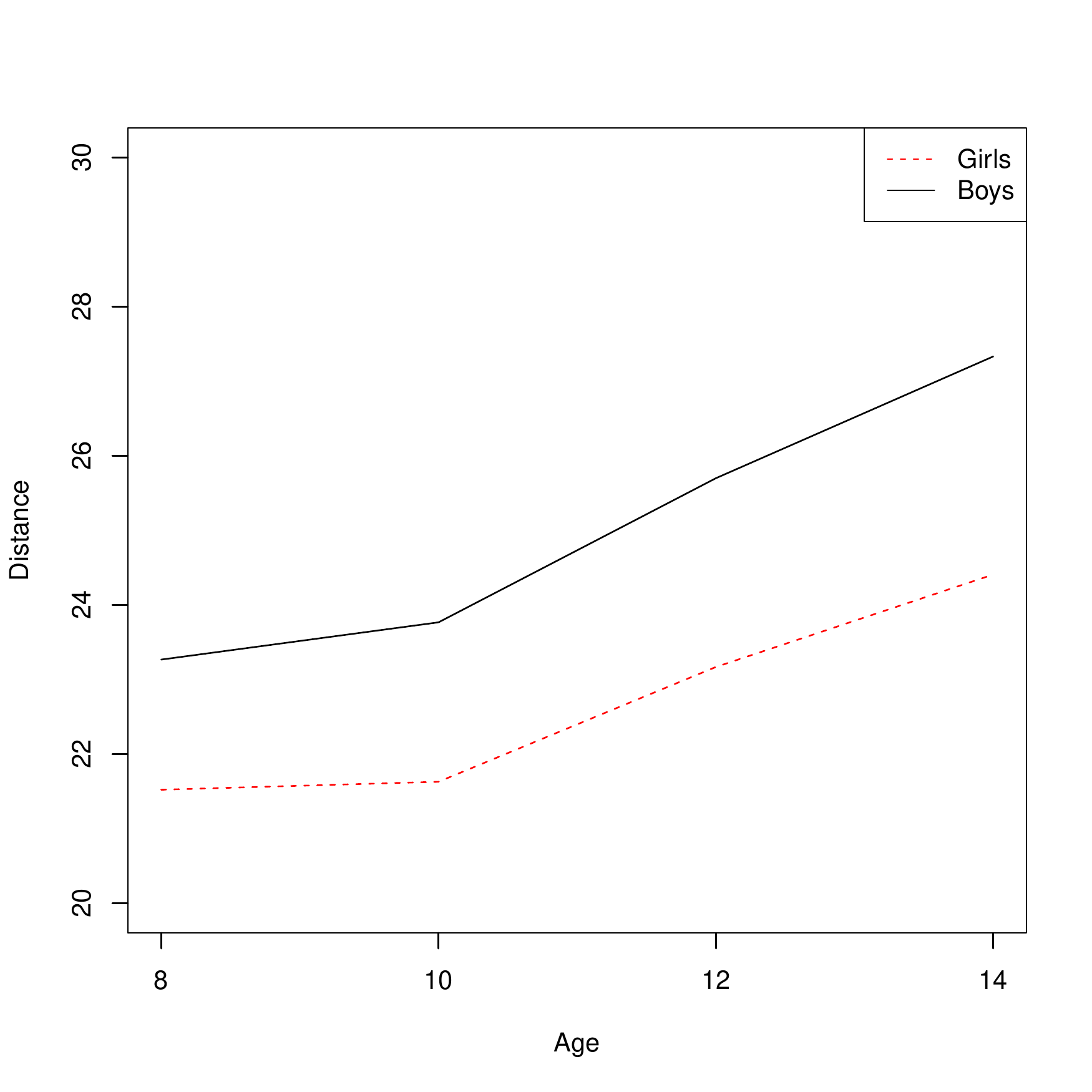}
\hfill
\includegraphics[width=2.8in]{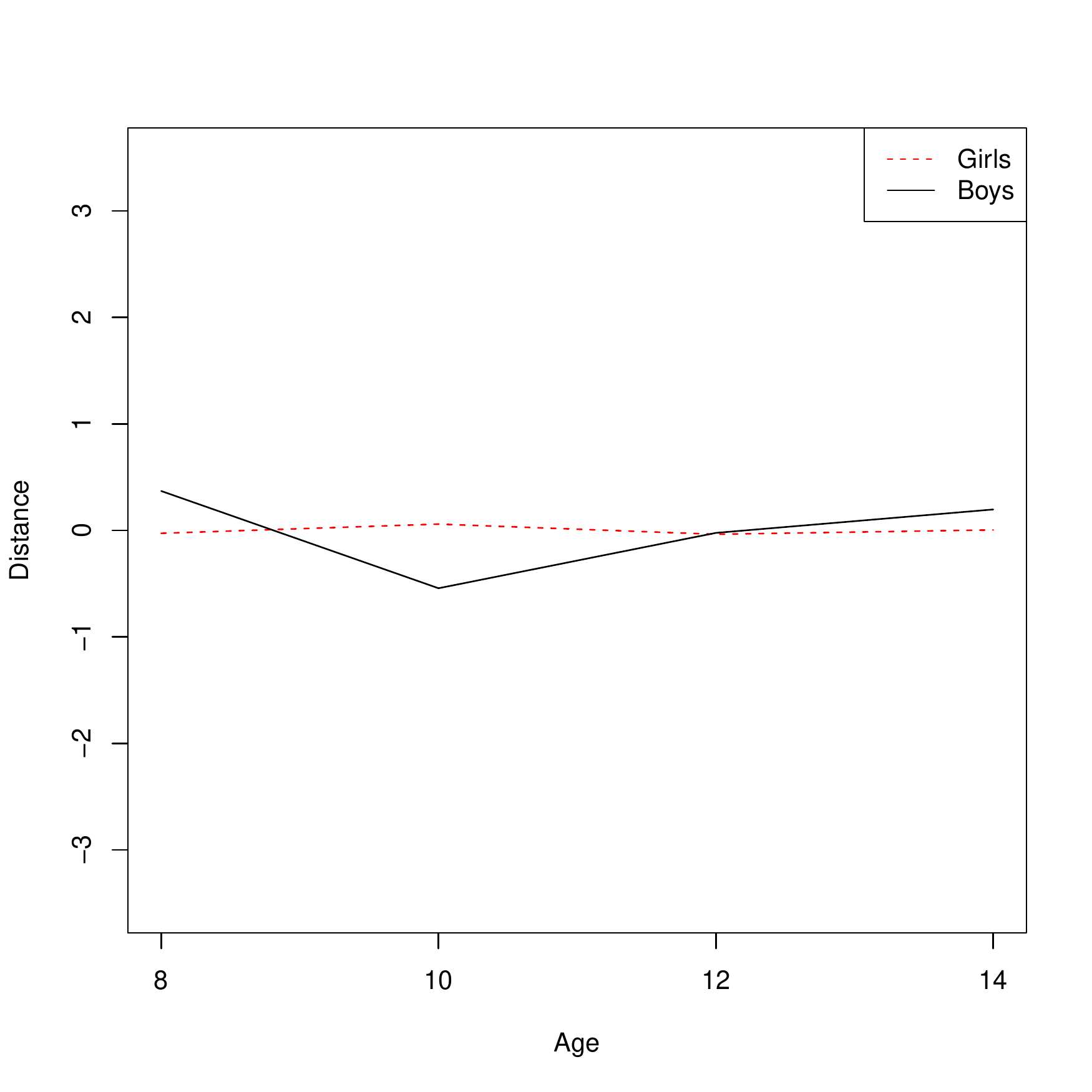}
\hfill}
\centerline{\hfill\makebox[2.1in]{c. Fitted values $\Yhat_{\Cm}$ from (\ref{igmlm})}
\hfill\makebox[2.1in]{d. Projected means $\Q_{\U}\Ybar$}\hfill}
\caption{\label{fig:dental}
Profile plots by sex of (a) the fitted vectors from envelope model (\ref{envmlm}), (b) means projected onto $\spn^{\perp}(\Gammabfhat)$, (c) the fitted vectors from the constrained model (\ref{igmlm}) and (d) means projected onto $\uspc^{\perp}$. The vertical axis for each plot is the distance for the plotted vectors.
}
\end{figure}

Figure~\ref{fig:dental}a gives a profile plot of the fitted vectors from envelope model (\ref{envmlm}).  The implied fit is quite good and close to the profile plot of the raw mean vectors shown in Supplement Figure~\ref{supfig:dental}a.  (Profile plots of residuals are also shown in Figure~\ref{supfig:dental}). Under envelope theory, the distribution of $\Q_{\Gammabfs}\Y$ should be independent of the predictor values, in this case sex.  The profile plot of $\Q_{\Gammabfhats}\Ybar$ by sex shown in Figure~\ref{fig:dental}b reflects this property.   Figures~\ref{fig:dental}cd show the corresponding plots from the fit of the constrained model (\ref{igmlm}). The fit of the constrained model altered the shape of the profile for girls so that it more closely matches that for boys, which was not done by the fit of the envelope model.  This type of conformity is an intrinsic property of constrained model (\ref{igmlm}).  

If there is uncertainty about the containment $\bspc \subseteq \uspc$ needed for the constrained model then it may be desirable to base an analysis on envelope model (\ref{envmlm}).  Otherwise, the results in the last two rows of Table~\ref{tab:dental} indicate that enveloping in the constrained model (\ref{igmlm}) is the best option from among those considered. 

We also applied the scaled envelope estimator discussed in Section~\ref{sec:envbeta}.  The asymptotic variances of the elements of the corresponding estimator of $\betabf$ did not differ materially from those shown in Table~\ref{tab:dental} for $\betabfhat_{\Um}$  and $\betabfhat_{\Em}$.
Then, scaling offered no gains in this example.  This was rather as expected since good scale estimation generally requires large sample size.  

\subsection{The China Health and Nutrition Survey}\label{subsec: CHNS}
 The China Health and Nutrition Survey (CHNS) was designed to evaluate the effects of the health, nutrition and family planning policies 
 on the health and nutritional status of its population \citep{popkin2009}.  
 The survey used a multistage, random cluster process to draw samples of households in 15 provinces and municipal cities that vary substantially in geography, economic development, public resources, and health indicators. 
 In totals, 9 surveys were carried out between 1989 and 2011.
 We  included in  our analysis only the 1209 individuals that participated in all the 9 surveys, giving a total of $9 \times 1209 = 10,881$ records.  Five individuals were deleted for having unreasonable changes in weight or height.  For instance, one individual had a height of 65 cm in the seventh survey but a height of 160 cm in all other surveys.  The baseline predictors we considered include age at the first survey, binary indicators for gender and region (urban or rural),  and a six-level indicator for highest education levels obtained at the first survey.
 About 98.2\% of the individuals in the analysis were over 21 years old. Age at first survey, gender and region were fully observed but there were $28$ individuals with missing education levels at baseline. We imputed the missing values with the education level collected at the next available visit.  The response was the change in BMI from baseline at the 8 followup surveys.  In the $10,881$ records, there was a total of 371 values of either missing height or weight information needed to calculate BMI. We assumed that height and weight were missing at random and imputed them by carrying the last observation forward.

 We compared the estimated asymptotic variances of the unconstrained estimator $\betabfhat_{\Um}$, the envelope estimator $\betabfhat_{\Em}$ and the constrained estimator $\betabfhat_{\Cm}$ from model (\ref{igmlm})  using $\Ubf^{T} = (1,t,t^{2})$, where $t$ is the time in years from baseline.  We also included the envelope version of the constrained estimator $\betabfhat_{\Ecm}$, the scaled envelope estimator $\betabfhat_{\Sem}$ from \citet{CookSu2013scaled} and its constrained version $\betabfhat_{\Secm}$ corresponding to model (\ref{igmlm}).  We used version (\ref{igmlm}) of the constrained model because we were interested in profile contrasts rather than modeling profiles per se.  
 
 Since $\betabfhat_{(\cdot)} \in \real{9 \times 9}$, we report in columns 4--9 of Table \ref{tb: data_var} various location statistics computed over the estimated variances of the individual elements in $\betabfhat_{(\cdot)}$.  Using these summary statistics as the basis for comparison,  we see that the estimators fall into two clear groups. The unconstrained estimator does the worst, followed closely by the envelope estimator and the constrained estimator.  The three envelope estimators listed in the last three rows of the table  do noticeably better than the first three.  Our assessment based on just the variance summary statistics and taking computational difficulty into account leads us to prefer the envelope constrained estimator $\betabfhat_{\Ecm}$.  The model order determined by BIC given in the third column of Table \ref{tb: data_var} tells a similar story.  Based on the actual BIC values, the unconstrained estimator in the first row appears clearly inferior to the others, while the scaled envelope model in the last row is clearly the best.  The remaining models are relatively difficult to distinguish.   We next give a few additional details.  
\begin{table}[ht]
\centering
\caption{BIC order, minimum, maximum, mean and quartiles $Q_{1}$--$Q_{3}$ of the estimated asymptotic variances of the elements in  $\betabfhat_{(\cdot)}$  for the CHNS study}\label{tb: data_var}
\medskip
\begin{tabular}{ccccccccc}
  \hline
 Estimator &Envlp. dim. &BIC order & Min & $Q_{1}$ & $Q_{2}$ & Mean & $Q_{3}$ & Max \\ 
  \hline
$\betabfhat_{\Um}$ &8 &6& 0.03 & 0.07 & 0.12 & 0.12 & 0.15 & 0.20 \\ 
 $\betabfhat_{\Em}$ & 2 &5& 0.02& 0.05 & 0.11 & 0.11 & 0.15 & 0.21 \\
 $\betabfhat_{\Cm}$ & 3 &4& 0.03 & 0.05 & 0.10 & 0.10 & 0.13 & 0.19 \\ 
 $\betabfhat_{\Ecm}$  &1 &3&  0.00 & 0.00 & 0.02 & 0.05 & 0.06 & 0.17 \\ 
 $\betabfhat_{\Secm}$ &1 &2&  0.00& 0.03 & 0.03 & 0.04 & 0.04 & 0.08 \\ 
 $\betabfhat_{\Sem}$  & 1 &1&  0.00 & 0.00 & 0.02 & 0.05 & 0.06 & 0.17 \\ \hline

\end{tabular}
\end{table}

The estimated dimensions of the various envelopes based using BIC are listed in the second column of Table~\ref{tb: data_var}.  We listed the maximum envelope dimension for the two non-envelope methods.  The variance gains for the envelope model over the unconstrained model shown in Table~\ref{tb: data_var} are reflected by the  two eigenvalues $(17.44, 15.90)$ of $\Omegabfhat$ and the six eigenvalues of $\Omegabfhat_{0}$ which ranged between $1.11$ and $1.62$.  Turning to the envelope version of constrained model (\ref{igmlm}), the estimated dimension of $\espc_{\Sigmabfs_{D|S}}(\aspc)$ using BIC was 1.  The variance gain over the unconstrained model shown in Table~\ref{tb: data_var} is again reflected by the value of $\Omegabfhat = 3 e^{-5}$ and the two eigenvalues of $\Omegabfhat_{0}$, $0.16$ and $2.74$.  As with the regular envelope model, the major variability lies in the immaterial part of the response. 

{\color{black}{
\subsection{Postbiotics study}

The aim of the {\color{blue} postbiotics} study \citep{Dunand} was to determine the protective capacity against Salmonella infection in
mice of the cell-free fraction (postbiotic) of fermented milk produced at
laboratory and industrial levels. The
capacity of the postbiotics produced by pH-controlled fermentation was evaluated to
stimulate the production of secretory IgA in {\color{blue} feces} and to protect mice against
Salmonella infection. There were  3 study groups with seven mice per group:  (i)
a control group (C), where mice received the unfermented
milk supernatant; (ii) an F36 group (F36), where mice 
received the cell-free supernatant obtained by DSM-100H
fermentation in 10\% (w/v) skim milk produced in the
laboratory; and (iii) an F36D group (F36D), where mice
received the product F36 diluted 1/10 in tap water.  {\color{blue} Feces samples  of }approximately 50 mg per mouse were collected once a week  for 6 weeks and the concentration of
secretory IgA (S-IgA) by ELISA was determinate. The response was the IgA measured over the 6 weeks period and
the predictors were the group indicators. 

\begin{figure}[ht!]
    \centering
          \includegraphics[width=3in]{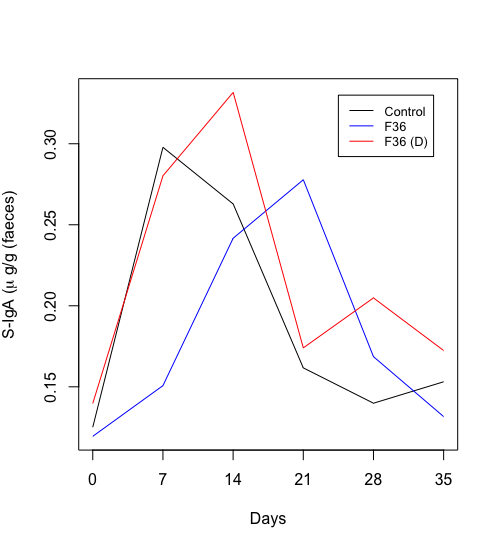}
        \caption{Average of IgA by group over time in the Posbiotics Study data}
        \label{1}
\end{figure}

The research question was whether there were differences of the IgA measures among the treatment groups. We
present the average response by group over the weeks in Figure \ref{1}. We set the control group as the baseline and therefore $\betabf \in \real{6 \times 2}$.
We calculate all estimators {\color{blue} based on envelope model} (\ref{igmlm}) because we were interested in profile contrasts rather than modeling profiles. We use $\Ubf^{T}(t) = (1,t/6,(t/6)^{2}, \cos ( 2 \pi t/6), \sin (2 \pi t/6))$, 
where $t=1,\dots, 6 $ are the weeks where the measures were taking.  
The unconstrained estimator $\betahat_{\Um}$ was considered in \citet{Dunand} and it did not show a difference between treatment groups, even when exploratory  {\color{blue} difference seem apparent from Figure \ref{1}.}


Table \ref{tb: data_var1} shows BIC, envelope dimension and MSE of the estimators. We listed the maximum envelope dimension for the two non-envelope methods as their estimated envelope dimensions.  The unconstrained estimator performs the worst and the scaled constrained envelope estimator performs the best in terms of both BIC and  efficiency.
\begin{table}[ht]
\centering
\caption{Envelope dimension, BIC, BIC order, and MSE   for the  Postbiotics Study}\label{tb: data_var1}
\medskip
\begin{tabular}{ccccc}
  \hline
 Estimator &Envlp. dim.& BIC  &BIC order &       MSE \\ 
  \hline
$\betabfhat_{\Um}$ &5 &-133.90 &6&   0.15  \\ 
 $\betabfhat_{\Em}$ & 1 &-163.52& 2&    0.13\\ 
 $\betabfhat_{\Cm}$ & 2 & -144.37   & 5& 0.15\\ 
 $\betabfhat_{\Ecm}$  &1 & -160.76 & 3&   0.14 \\ 
 $\betabfhat_{\Sem}$  & 1 &  -152.22  & 4&   0.13 \\ 
 $\betabfhat_{\Secm}$ &1 &-251.48   & 1&  0.13\\ \hline
\end{tabular}
\end{table}

To address the researcher question, we {\color{blue} used} the $p$-values of the $\betahat$ components. From Table \ref{tb: data_var2} we can see that the unconstrained estimator does not reveal any differences, which aligns with the findings in  \citet{Dunand}. None of the estimators demonstrate any evidence of difference between F36D group and the control group at any time. On the other hand, $\betahat_{\Secm}$ reveals a significance difference between the control and  {F36} groups in all followup weeks. The $p$-values for such a comparison of $\betahat_{\Em} $ are {\color{blue} clearly significant only in week 3}. Other estimators also fail to find all followup weeks significant between F36 and control groups, e.g., the scaled envelope is not significant in week 5 and 6, and constrained envelope is significant only in week 2. 

 \begin{table}[ht]
\centering
\caption{The $p$-values for coefficients for $\betahat_{\Um}$, $\betahat_{\Em}$ and $\betahat_{\Secm}$ }\label{tb: data_var2}
\medskip
\begin{tabular}{cccccccc}
  \hline
  week&\multicolumn{3}{c}{F36 vs control}&& \multicolumn{3}{c}{F36 D vs control}\\
\cmidrule{2-4}\cmidrule{6-8}
&$\betahat_{\Um} $ & $\betahat_{\Em} $   & $\betahat_{\Secm} $ && $\betahat_{\Um} $ & $\betahat_{\Em} $  &$\betahat_{\Secm} $  \\ 
  \hline
1& 0.91 & 0.07&0.13&& 0.77 &0.27& 0.30\\
 2& 0.09 &0.10& 0.01 &&0.83 &0.28&0.21 \\
 3& 0.83& 0.01&  0.01 && 0.48 &0.20&0.22 \\
 4& 0.26 &0.06&0.02 &&  0.90 &0.23&0.22 \\
  5&0.55 &0.05&0.00 &&  0.16 & 0.20&0.20\\
  6&0.57 &0.63&  0.01 && 0.59 &0.64& 0.21\\
 \hline
\end{tabular}
\end{table}

   The variance gains for the scale version of the constrained envelope model over the unconstrained model  (and therefore the $p$-values) are reflected by the   eigenvalue  $1e^{-4}$ of $\Omegabfhat$ and the four eigenvalues of $\Omegabfhat_{0}$ which are $23.06$, $13.67$,   $0.41 $ and $  0.22$.  The reason for the envelope estimator to be not as significant when comparing F36 and control groups is that
there is not as big a discrepancy between the eigenvalues of $\Omegabfhat $ ($2e^{-3} $)  and the eigenvalues of $\Omegabfhat_0$ ($ 0.02$, $0.04$,  $0.03$, $0.01$, and $4e^{-3}$).  

}}

 \section{Discussion}
  
 The primary computational step for all of the envelope methods described herein involves finding $\Ghat = \arg \min_{\G \in \gspc} \log|\G^{T} \M_{1}\G| + \log | \G^{T} \M_{2} \G|$ over a class $\gspc$ of semi-orthogonal matrices, where the inner product matrices $\M_{1}$ and $\M_{2}$ depend on the application.  The R package Renvlp by M. Lee and Z. Su contains a routine for minimizing objective functions of this form.  Computations are straightforward once $\Ghat$ has been found.   Renvlp also implements specialized methodology for data analysis under envelope model (\ref{envmlm}) and partial envelope model.  The associated routines can be modified for the models described herein.  Description of and links to packages for envelope methods are available at z.umn.edu/envelopes.
 
 We relegated discussion of certain well-established aspects of envelope methodology to the Supplement.  Non-normality and the bootstrap are discussed in Section~\ref{sec:boot} and methods for selecting the envelope dimension are reviewed in Section~\ref{sec:estu}.   Enveloping for $(\alphabf_0,\alphabf) $ jointly is discuss in Section \ref{both} and finally a brief discussions of envelopes and Rao's simple structure is in Section~\ref{sec:structure}
 Extensions to unbalanced data and random effects models requiere additional research,


%

\renewcommand{\refname}{References}
\bibliographystyle{apalike}
\bibliography{EnvGC}{}

\end{document}